\documentclass[letterpaper,12pt]{article}
\usepackage[top=1in, bottom=1in, left=1.25in, right=1.25in]{geometry}  

\usepackage{setspace} 
\usepackage{sectsty}
\allsectionsfont{\singlespacing}

\usepackage[compact]{titlesec}
\titlespacing{\section}{0pt}{*0}{*0}
\titlespacing{\subsection}{0pt}{*0}{*0}
\titlespacing{\subsubsection}{0pt}{*0}{*0}

\usepackage{caption}
\usepackage{etoolbox}
\BeforeBeginEnvironment{equation*}{\begin{singlespace}\vspace*{-\baselineskip}}
\AfterEndEnvironment{equation*}{\end{singlespace}\noindent\ignorespaces}
\BeforeBeginEnvironment{equation}{\begin{singlespace}\vspace*{-\baselineskip}}
\AfterEndEnvironment{equation}{\end{singlespace}\noindent\ignorespaces}
\BeforeBeginEnvironment{align}{\begin{singlespace}\vspace*{-\baselineskip}}
\AfterEndEnvironment{align}{\end{singlespace}\noindent\ignorespaces}
\BeforeBeginEnvironment{align*}{\begin{singlespace}\vspace*{-\baselineskip}}
\AfterEndEnvironment{align*}{\end{singlespace}\noindent\ignorespaces}
\BeforeBeginEnvironment{eqnarray}{\begin{singlespace}\vspace*{-\baselineskip}}
\AfterEndEnvironment{eqnarray}{\end{singlespace}\noindent\ignorespaces}
\BeforeBeginEnvironment{eqnarray*}{\begin{singlespace}\vspace*{-\baselineskip}}
\AfterEndEnvironment{eqnarray*}{\end{singlespace}\noindent\ignorespaces}

\usepackage{authblk} 

\usepackage{amsmath, amsthm}  
\usepackage{mathtools}   
\usepackage{bm}
\usepackage{amsfonts}

\usepackage{graphicx} 
\usepackage{enumerate}
\usepackage[authoryear]{natbib}  
\usepackage{url} 

\usepackage{enumitem}   
\usepackage[table,xcdraw,dvipsnames]{xcolor}   
\usepackage{pdflscape}
\usepackage{afterpage}
\usepackage{capt-of}
\usepackage{adjustbox}

\usepackage{multirow}   
\usepackage{tabularx} 

\usepackage{float}   
\usepackage{nameref}   

\usepackage[font=normalsize]{caption}  
\usepackage{subcaption}  
\usepackage[titletoc,title]{appendix}  
\usepackage{longtable} 

\usepackage[final]{hyperref} 
\hypersetup{
	colorlinks=true,       
	linkcolor=blue,        
	citecolor=blue,        
	filecolor=magenta,     
	urlcolor=blue         
}

\graphicspath{ {images/} }

\newcommand{\utwi}[1]{\mbox{\boldmath $ #1$}}

\newcommand{\bone}{{\utwi{1}}}

\newcommand{\bA}{{\utwi{A}}}

\newcommand{\bE}{{\utwi{E}}}
\newcommand{\bF}{{\utwi{F}}}

\newcommand{\bH}{{\utwi{H}}}
\newcommand{\bI}{{\utwi{I}}}

\newcommand{\bM}{{\utwi{M}}}

\newcommand{\bQ}{{\utwi{Q}}}

\newcommand{\bW}{{\utwi{W}}}
\newcommand{\bX}{{\utwi{X}}}

\newcommand{\bZ}{{\utwi{Z}}}

\newcommand{\bOmega}{{\utwi{\Omega}}}

\newcommand{\calM}{{\mathcal M}}

\newtheorem{remarkx}{Remark}

\newtheorem{condx}{Condition}

\DeclarePairedDelimiterX{\norm}[1]{\lVert}{\rVert}{#1}   

\newcommand{\E}[1]{{\rm E} \left \{ #1 \right \}}    

\newcounter{question}

\renewcommand{\hat}{\widehat}

\numberwithin{equation}{section}  

\newcounter{CondCounter}

\newcommand{\blind}{0}

\begin{document}
\if0\blind
{   \singlespacing
	\title{\bf \Large Modeling Dynamic Transport Network with \\ Matrix Factor Models: with \\ an Application to International Trade Flow\footnote{ Supported in part by NSF Grants DMS-1503409, DMS-1737857, DMS-1803241 and IIS-1741390. Corresponding author: Rong Chen, email rongchen@stat.rutgers.edu}}
	\author[1]{Elynn Y. Chen}
	\author[2]{Rong Chen}
	\affil[1]{Department of Operations Research and Financial
Engineering, Princeton University}
	\affil[2]{Department of Statistics and Biostatistics, Rutgers University}
	\date{\vspace{-5ex}}
	\maketitle
} \fi

\if1\blind
{   \singlespacing
	\bigskip
	\bigskip
	\bigskip
	\title{\bf Modeling Dynamic Transport Network with \\ Matrix Factor Models: with \\ an Application to International Trade Flow}
	\author{}
	\date{\vspace{-5ex}}
	\maketitle
	\medskip
} \fi

\bigskip

\begin{abstract}
International trade research plays an important role to inform trade policy and shed light on wider issues relating to poverty, development, migration, productivity, and economy. With recent advances in information technology, global and regional agencies distribute an enormous amount of internationally comparable trading data among a large number of countries over time, providing a gold mine for empirical analysis of international trade. Meanwhile, an array of new statistical methods are recently developed for dynamic network analysis. However, these advanced methods have not been utilized for analyzing such massive dynamic cross-country trading data. 
International trade data can be viewed as a dynamic transport network because it emphasizes the amount of goods moving across a network. 
Most literature on dynamic network analysis concentrates on the connectivity network that focuses on link formation or deformation rather than the transport moving across the network. 
We take a different perspective from the pervasive node-and-edge level modeling: the dynamic transport network is modeled as a time series of relational matrices. 
We adopt a matrix factor model of \cite{wang2018factor}, with a specific interpretation for the dynamic transport network. Under the model,
the observed surface network is assumed to be driven by a
latent dynamic transport network with lower dimensions. The proposed method is able to unveil the latent dynamic structure and achieve the objective of dimension reduction.
We applied the proposed framework and methodology to a data set of monthly trading volumes among 24 countries and regions from 1982 to 2015. Our findings shed light on trading hubs, centrality, trends and patterns of international trade and show matching change points to trading policies. The dataset also provides a fertile ground for future research on international trade.
\vspace{1em}

\end{abstract}

\section{Introduction}


International trade research addresses the important questions about the drivers and effects of international trade in goods and services, as well as the design and implications of trade policy, regional integration and the global trading system. It provides vital information for trade and economy policy making and, at the same time, sheds light on wider issues relating to poverty, development, migration, productivity and global economy. Traditionally, international trade research emphasizes more on the theoretical aspect \citep{sen2010international}. An empirical shift of the discipline did not occur until the beginning of this century \citep{davis2001whatrole}.

With recent advances in information technology, the United Nation system (UN), the World Bank, the International Monetary Fund, among others, collect, produce and distribute an enormous amount of internationally comparable trading data over time, providing a gold mine for empirical analysis of international trade. These dynamic network data provide a wide variety of information (e.g. the patterns of interactions, the evolution of the relative importance, and the natural grouping of actors in the network) that can be extracted to understand many aspects of international trade. \cite{smith1992structure} measure the structure of world economic system and identify the roles that particular countries play in the global division of labor by using block modeling \citep{lorrain1971structural} on international commodity trade flows at three time points (year 1965, 1970 and 1980).  \cite{kim2002longitudinal} study globalization and regionalization in international trade by calculating network related quantities, such as in-degree, out-degree, centralization and block densities, during three consecutive periods. \cite{mahutga2006persistence} examines the structural equivalence in international trade by conducting correspondence analysis -- one of a family of techniques based on the Singular Value Decomposition -- to the equivalence matrix that is constructed to summarize the degree of regular equivalence for each pair of countries from the network data. See also \cite{hafner2009network} for a survey of network analysis in international relations. The econometric tools necessary for the empirical analysis of such data that reflect pairwise interactions between economic agents are still in their infancy. Moreover, much of the existing empirical trade literature is concerned with patterns of international trade at a point of time. This focus of empirical work stands in marked contrast with the theoretical literature on growth and trade that are dynamic and evolving over time.

In this paper, we propose an empirical framework for analyzing the evolution of patterns of international trade over time. We model the trade flow data as time series of square matrices that describe pairwise relationships among a set of entities. Specifically, trade data between $n$ countries over a period of time can be represented as a matrix-variate time series $\{ \bX_t \}_{t=1:T}$, where $\bX_t$ is a $n \times n$ matrix, and each element $x_{ij,t}$ is the directed volume of trade from country $i$ to country $j$ at time $t$. The $i$-th row represents data for which country $i$ is the exporter and the column $j$ represents data for which country $j$ is the importer. We explore the underlying latent lower-dimensional structure of the dynamic network by using variations of the matrix factor model \citep{wang2018factor}. The latent networks and their connection to the surface networks provides a clear view of the evolution of international trade over three decades. The resulting lower dimensional representation of the dynamic network can be used for second-step analyses such as prediction of matrix time series.

Researchers have studied dynamic network/relational data analysis from various aspects. Snijders and colleagues \citep{snijders2001statistical, huisman2003statistical,snijders2005models,snijders2006statistical,snijders2007modeling, snijders2010maximum, snijders2010introduction} developed an actor-driven, or actor-oriented, model for network evolution that incorporates individual level attributes. The change of network structure is the result of the economic rational choice of social actors (selection) and the characteristics of others to whom they are tied (influence). They apply the analysis to an evolving friendship network and the focus is link evolution between friends. \cite{hanneke2010discrete} and \cite{krivitsky2014separable} introduced a class of temporal exponential random graph models for longitudinal network data (i.e. the networks are observed in panels). They model the formation and dissolution of edges in a separable fashion, assuming an exponential family model for the transition probability from a network at time $t$ to a network at time $t+1$. \cite{westveld2011mixed} represent the network and temporal dependencies with a random effects model, resulting in a stochastic process defined by a set of stationary covariance matrices. \cite{xing2010state} extends an earlier work on a mixed membership stochastic block model for static network \citep{airoldi2008mixed} to the dynamic scenario by using a state-space model where the mixed membership is characterized through the observation function and the dynamics of the latent `tomographic' states are defined by the state function. Estimation is based on the maximum likelihood principal using a variational EM algorithm. These methods focus on the connectivity of the nodes, that is, 0-1 status rather than the weights of the links. The methods are deduced from random graph theory and model the relational data at relation (edge) or entity (node) level, and thus often confronted with computational challenges, over-parameterization, and over-fitting issues.

In contrast to the pre-existing research in dynamic network analysis, the approach we propose focuses more on the edges (traffic flows) of the network and their dynamic properties. The nodes are characterized by the flows to/from other nodes. Specifically, the traffic flows in a network are represented as a time series of matrix observations -- the relational matrices-- instead of the traditional nodes and edges characterization. The structure of a matrix preserves the pair-wise relationships and the sequence of matrices preserves the dynamic property of such relationships. We adopt a matrix factor model where the observed surface dynamic network is assumed to be driven by a latent dynamic network with lower dimensions. The linear relationship between the surface network and the latent network is characterized by unknown but deterministic loading matrices. The latent network and the corresponding loadings are estimated via an eigenanalysis of a positive definite matrix constructed from the auto-cross-moments of the network times series, thus capturing the dynamics presenting in the network. Since the dimension of the latent network is typically small or at least much smaller than the surface network, the proposed model often yields a concise description of the whole network series, achieving the objective of dimension reduction. The resulting latent network of much smaller dimensions can also be used for downstream microscope analysis of the dynamic network.

Different from \cite{xing2010state} that summarize the relational data by the relationships between a small number of groups, we impose neither any distributional assumptions on the underlying network nor any parametric forms on its moment function. The latent network is learned directly from the data with little subjective input. The meaning of the nodes of the latent network in our model is automatically learned from the data and is not confined to the `groups' to which the actors belong, which provide a more flexible interpretation of the data. Additionally, our modeling framework is very flexible and extendable: using a matrix factor model framework, it can accommodate continuous and ordinal relational data. It can be extended to incorporate prior information on the network structure or include exogenous and endogenous covariate as explanatory variables of the relationships. Although the focus of the analysis in this paper is to estimate the latent lower-dimensional network underlying the surface network, the innovative idea of modeling dynamic network as time series of relational matrices is simple, yet quite general. Autoregressive models for matrix-variate time series \citep{chen2018autoregressive} can be included under this framework to model the dynamics of latent matrix factors and to provide predictions of the network flows.

The remaining part of the paper is organized as follows. In Section
\ref{sec:dataEDA}, we describe the international trade flow data from 1981 to 2015 and
present some explanatory analysis results. In Section \ref{sec:model}, we introduce two factor models for network time series data and discuss their interpretations. In Section \ref{sec:est}, we present the estimation procedure and the properties of estimators. In Section \ref{sec:application}, we apply the proposed factor models to the international trade data described in Section \ref{sec:dataEDA}.
 In Section \ref{sec:summary}, we summarize this paper and present future research directions.

\section{International Trade Flow Data and Exploratory Analysis} \label{sec:dataEDA}
\subsection{Trade Flow Data}
In the following analysis, we use monthly multilateral import and export volumes of commodity goods among 24 countries and regions over the 1982 -- 2015 period. The data come from the International Monetary Fund (IMF) \emph{Direction of Trade Statistics} (DOTS) \citep{IMF-DOTS}, which provides monthly data on countries' exports and imports by their partners. The source has been widely used in international trade analysis such as the Bloomberg Trade Flow.
Even though IMF-DOTS provides data from January, 1948 containing 236 countries, the quality of data varies across time and countries.
Some countries failed to report their volumes of trade in some or all years.
Most of these missing cases are concentrated in small and underdeveloped countries or current or former Communist countries.
In this study, we restrict the sample to 24 countries and regions from three major trading groups, namely NAFTA, EU and APEC, over a 408-month period from January, 1982 to December, 2015. The countries and regions used in alphabetic order are Australia, Canada, China Mainland, Denmark, Finland, France, Germany, Hong Kong, Indonesia, Ireland, Italy, Japan, Korea, Malaysia, Mexico, Netherlands, New Zealand, Singapore, Spain, Sweden, Taiwan, Thailand, United Kingdom, and United States.

We use the import CIF data of all goods denominated in U.S. dollars since it is generally believed that they are more accurate than the
export ones \citep{durand1953countryclassification, linnemann1966econometric}. This is especially true when we are interested in tracing countries of production and consumption rather than countries of consignment or of purchase and sale \citep{linnemann1966econometric}. The figures for exports are determined by imputing them from imports. For example, Canada's export volume to France is determined as France's import volume from Canada. This calculation is done to make world total imports and exports equal. Note that the trade data for Taiwan as
a reporting
 region
is not published in the IMF-DOTS. In this paper, import data for Taiwan are imputed from the export data reported by its partner countries. As \cite{linnemann1966econometric} notes, in order to reduce the effect of incidental transactions of unusual size and of incidental difficulties in trade contract, trade flows were measured as three-month averages, rather than as direct observations of a particular month. For example, the trade flows in March, 2014 are the averages of those in February, March, and April of 2014.

\subsection{Exploratory Analysis}

The dynamic trading network can be cast into a time series of relational matrices that record the ties (trading volumes) between the nodes (countries) in the network.
The length of our network matrix time series is 408 months.
At each time point, the observation is a square matrix whose rows and columns represent the same set of $24$ countries. Each row (column) corresponds to an export (import) country. Each cell in the matrix contains the dollar trading volume that the exporting country exports to the importing country. The diagonal elements are undefined.

\begin{figure}[htbp!]
	\centering
	\includegraphics[width=\textwidth, keepaspectratio]{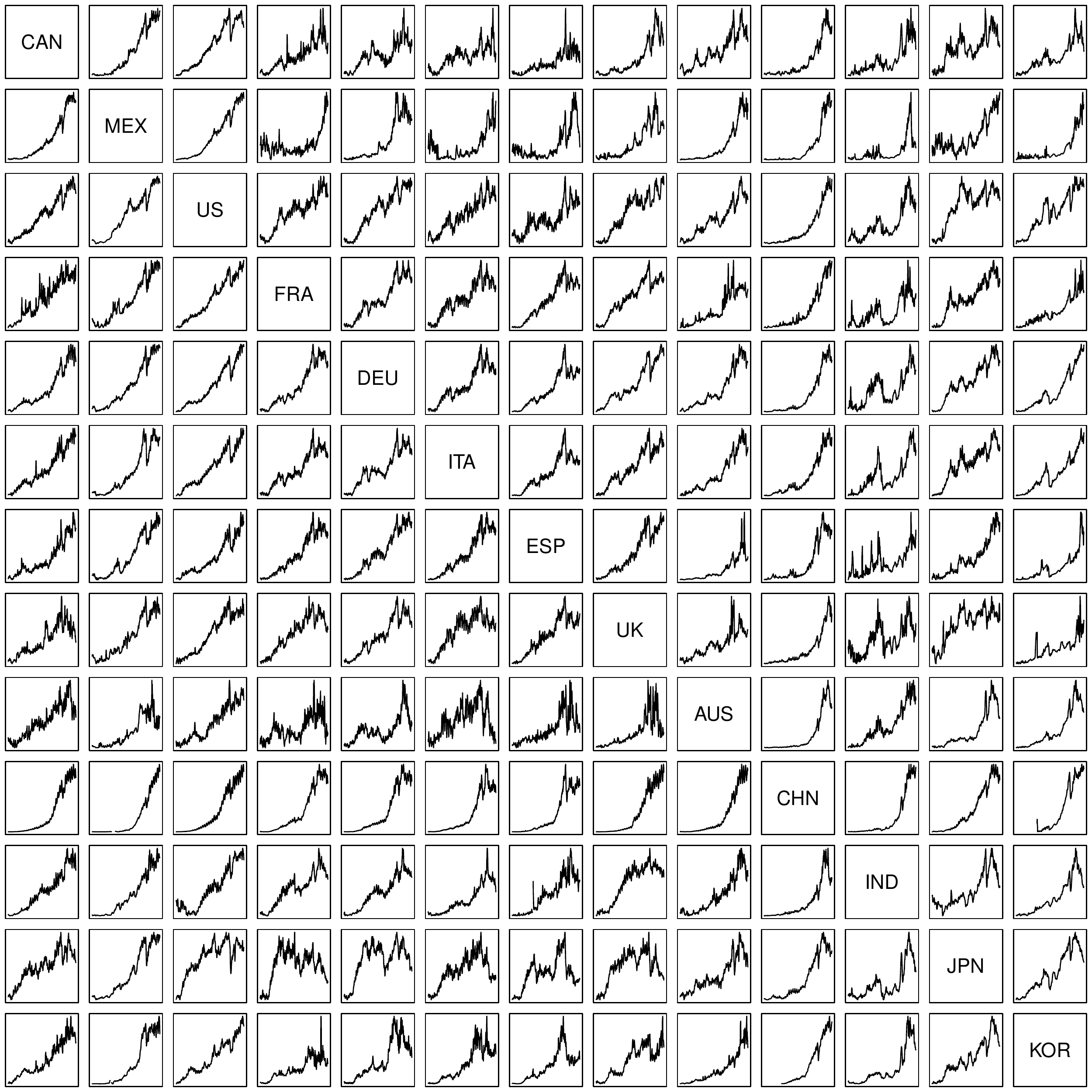}
	\caption{Time series plots of the value of good traded among 13 countries over 1982 -- 2015. The plots only show the patterns of the time series while the amplitudes are not comparable between plots because the range of the y-axis are not the same. }
	\label{fig:mts_13}
\end{figure}

Figure \ref{fig:mts_13} plots the time series of
trading volumes in U.S. dollar among top 13 countries from January, 1982 to December, 2015 in our dataset. Each time series is normalized for ease of visualization. These 13 countries are representative of all countries and regions in our dataset. They falls into three major groups: Canada, Mexican, and United States compose the NAFTA group; France, Germany, Italy, Spain, and United Kingdom are in the EU group; Australia, China Mainland, Indian, Japan and Korea belong to the APEC group. Overall, all countries experienced rapid growth in trades along with the accelerating wave of globalization. The world saw largest collapse in the value of good traded in 2009 when the impact of the global financial crisis was at its worst. Some actually have not recovered yet.
For example, we see that Spain's downturn in import has not recovered so far, though
its export has mostly recovered. While the upward trends are shared among all countries, the pattern of trading are more alike among countries within the same group. For example, the exports time series of the five European countries resembles more to each other than to the exports time series of the Asian countries.

\begin{figure}[htbp!]
	\centering
	\includegraphics[width=\textwidth, keepaspectratio]{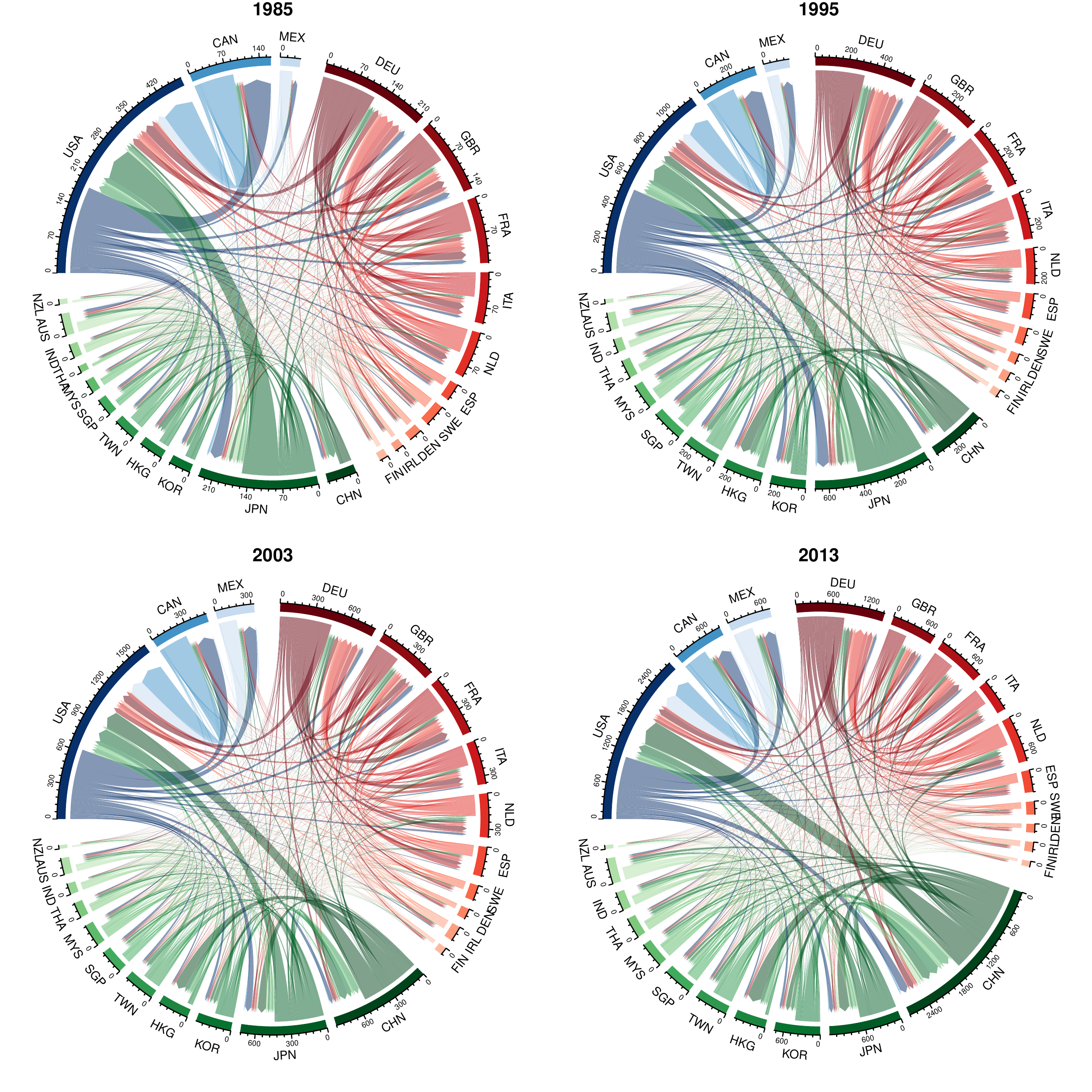}
	\caption{Circular trading plots that are representative of the bilateral relationship patterns in the 1980's, 1990's, 2000's and 2010's. The arrowhead indicates the direction of exports. The width of the arrow at its base represents the size of trade flow. Numbers on the outer section axis correspond to the size of trading flows in billion dollars.  }
	\label{fig:chordDiag_4}
\end{figure}

In order to illustrate the pattern of bilateral relationships, a set of four circular trading plots are shown in Figure \ref{fig:chordDiag_4}. The direction of flow is indicated by the arrowhead. The size of the flow is shown by the width of the arrow at its base. Numbers on the outer section axis, used to read the size of trading flows, are in billions. Each plot is based on the monthly flows over a one-year period, aggregated to selected annual volumes. Note that the four plots are representative of the bilateral relationship patterns in the 1980's, 1990's, 2000's and 2010's. 

For the three groups (EU, NAFTA, and APEC), most of the trade flows occur within the same group. This phenomenon is most prominent within the EU group where the imports and exports are all in red shade that denotes EU countries in Figure \ref{fig:chordDiag_4}. The trade flows of NAFTA countries are least confined within the group, mainly because the U.S. alone trades a lot with both EU and APEC countries.

For individual countries, most noticeable are changes in the share and direction of trade of U.S., China Mainland, Mexico and Japan. Over the years, U.S. maintains the most distinctive one among all countries because of its large trading volumes and wide range of trading counter-parties. The destinations of U.S. exports gradually shift from Japan and European countries to China Mainland and Mexico. In the 1980's Japan accounted for the largest importing and exporting flow among APEC countries. As shown clearly in Figure \ref{fig:chordDiag_4}, China Mainland's slice of pie in global trades grew steadily in size and becomes the largest in the 2010's. Mexico experienced a similar steady growth in global trades although less prominent than that of China Mainland. The trading patterns are most stable of the EU countries. The EU countries almost keep the same portions in the size of imports and exports over years.

The explanatory statistical analysis and visualization tools provide very clear and powerful but only descriptive observations. It is clear that there exists a possibly lower dimensional latent network, underlying the large scale dynamic network on the surface. However, there are few statistical tool available to quantify this latent structure. In the next section, we present a new methodology that is able to quantify the latent dynamic networks that underpins the observed surface dynamic networks as well as the relationship that connect the latent networks and the surface networks.

\section{Matrix Factor Models for Dynamic Transport Network} \label{sec:model}

In this section, we propose a new general methodology for investigating the evolving structure of dynamic networks. Here we focus
on the traffic flows in the dynamic network such as international import-export trade network, air-passenger volume between cities, and the number of directional interactions among people. The networks in our current considerations are typically dense. We refer to such a dynamic network as dynamic transport network. In the proposed framework, the bilateral relationships in the network at time $t$ is recorded in a relational matrix $\bX_t$ whose rows and columns corresponds to the same set of actors in the network. The elements of $\bX_t$ record information of the ties between each pair of the actors. The dynamic features of the networks are characterized by the temporal dependencies among consequential observations. Specifically, the entire dynamic networks is modeled as a sequences of temporally dependent matrix-variate $\{ \bX_t \}_{1:T}$. An important attribute of this modeling framework is that it captures
both the network structure and the temporal dynamics of the dynamic networks at a high level without any distributional assumption, different from the most common node-and-edge level modeling.

To formalize the methods, let $\bX_t$ represent the $n$ by $n$ relational matrix of observed pairwise asymmetrical relationships at time $t$, $t = 1, \ldots, T$. A general entry of $\bX_t$, denoted as $x_{ij,t}$, represents the directed relationship of actor $i$ to actor $j$. For example, in international trade context $x_{ij,t}$ expresses the volume of trade flow from country $i$ to country $j$ at time $t$; in the transportation context $x_{ij,t}$ represents the volume, fare, or length of a trip from location $i$ to location $j$ starting at time $t$.

Our model for dynamic transport network can be written as:
\begin{equation} \label{eqn:fac_AA}
	\bX_t = \bA \bF_t \bA' + \bE_t,
\end{equation}
where $\bA$ is an $n \times r$ (vertical) matrix of "loadings" of the $n$ actors on a relatively few $r$ $( < n)$ components (we will call them ``hubs''). $\bF_t$ is a small, usually asymmetric, $r$ by $r$ matrix giving the directional relationships among the $r$ latent hubs, and $\bE_t$ is simply a matrix of error terms. Since $\bX_t$ does not have diagonal elements, $\bE_t$ has a missing diagonal as well. Loading matrix $\bA$ relates the observed actors to the latent hubs and $\bF_t$ describes the dynamic interrelations among the hubs.

The interpretation of Model~(\ref{eqn:fac_AA}) can be demonstrated by referring to an example of international trade. 
Model~(\ref{eqn:fac_AA}) describes $r$
basic factors underlying the pattern of international trade for a given set of countries. One can view the latent factors in Model~(\ref{eqn:fac_AA}) as hubs and the export-import trading among the countries all go through these hubs. Each country exports to the $r$
hubs in certain distributions (determined by the loading matrix $\bA$) and import from the hubs in the same distributions. The hubs trade,
on behave of the participating countries,
among themselves and also within the hubs. The trading volume among the hubs are reflected by the factor matrix $\bF_t$, which is changing over time (dynamic). The $(k,l)$-th element $f_{kl,t}$ reflects the export trading volume from hub $k$ to hub $l$ at time point $t$. By examine the loading (distribution) from each country, it is often possible to `label' the hubs, even though they are purely estimated from data, instead of through construction. For example, if a hub's import are mainly contributed by members of major energy (such as oil and gas) production countries, then it can be labeled as an energy
 hub. Or if a hub's contribution mainly comes from countries in a geographic region such as Euro Zone, then it can be labeled as Europe hub. Note that under Model~(\ref{eqn:fac_AA}),
\[
x_{ij,t} = \underset{k, l}{\sum} a_{i,k} \, f_{kl,t} \, a_{j,l} + \epsilon_{ij,t}.
\]
Each term $a_{i,k} \, f_{kl,t} \, a_{j,l}$ can be interpreted as the (export) contribution of country $i$ to hub $k$, and the (import) contribution of country $j$ to hub $l$ in the export activity from hub $k$ to $l$. The total volume $x_{ij,t}$ is the summation of the exporting volumes from country $i$ to $j$ through all the latent hubs.

An interesting feature of the above model is that, while $\bF_t$ is allowed to be asymmetric, the left and right loading matrices $\bA$ are required to be identical. This provides a description of data in terms of asymmetric relations among a \textit{single} set of hubs rather than envisioning a different set of hubs. For example, in our international trade example Model~(\ref{eqn:fac_AA}) implies that the countries have the same set of hubs in their ``exporting'' role as they have in their ``importing'' role. A second possible approach, where the left loading matrix may be different from the right one, can be written as:
\begin{equation} \label{eqn:fac_A1A2}
\bX_t = \bA_1 \bF_t \bA_2' + \bE_t,
\end{equation}
where $\bA_1$ ($\bA_2$) is the $n \times r_1$ ($n \times r_2$) vertical loading matrices of the $n$ row (column) actors on $r_1$ ($r_2$) $( < n)$ hubs.
Matrices $\bF_t$ ($r_1\times r_2$)
and $\bE_t$ are defined the same as in those in (\ref{eqn:fac_AA}). This formulation is the matrix factor model considered in \cite{wang2018factor}.

\begin{figure}[htbp!]
\centering
\begin{tabular}{@{}cc@{}}
  \includegraphics[width=.45\textwidth]{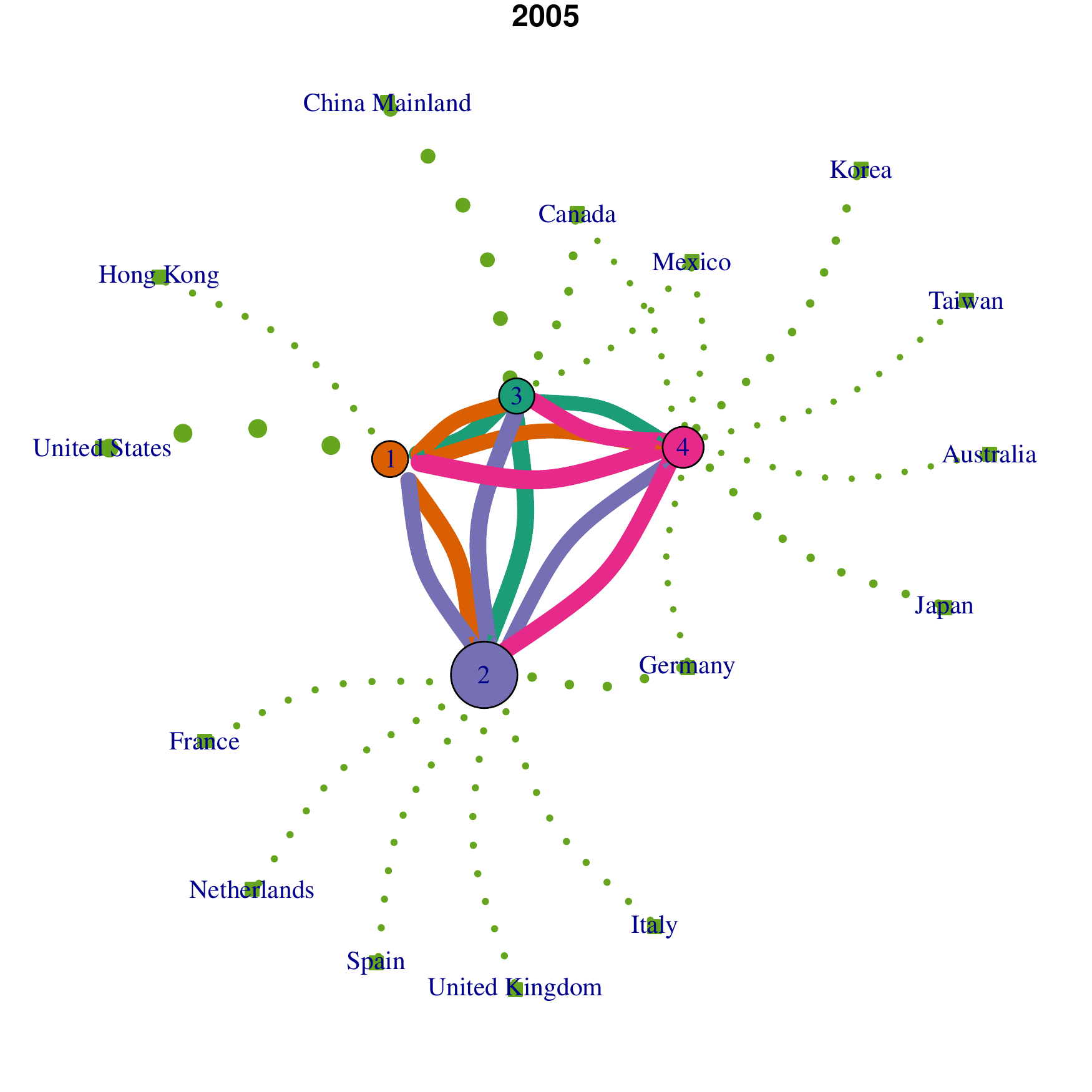} & \includegraphics[width=.5\textwidth]{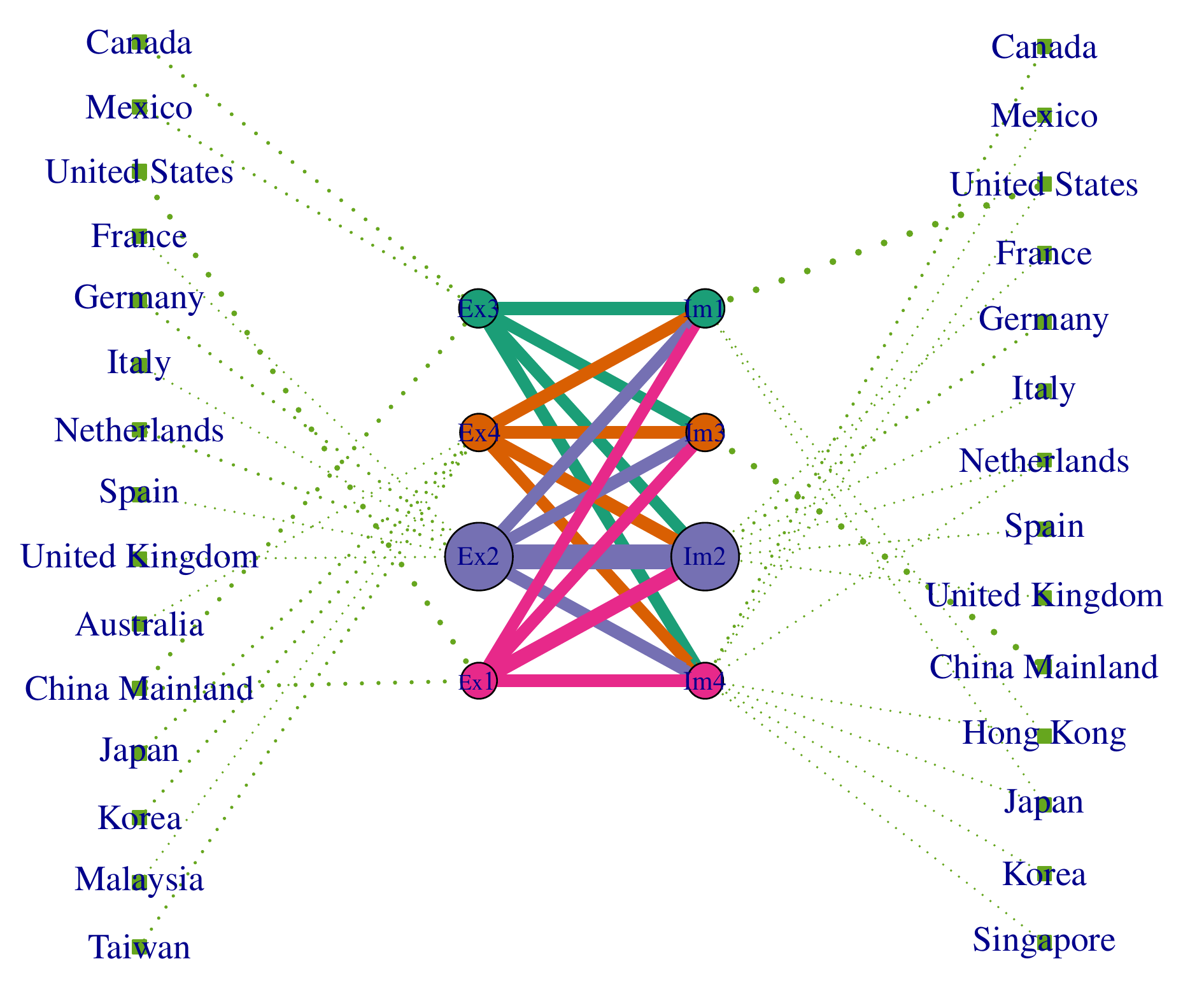}
\end{tabular}
\caption{Comparison between Model~(\ref{eqn:fac_AA}) (left) and Model~(\ref{eqn:fac_A1A2}) (right). }
\label{fig:AA_A1A2_network_compare}
\end{figure}

Model~(\ref{eqn:fac_AA}) describes asymmetric relationships among actors in terms of asymmetric relationships among a single set of underlying hubs. Model~(\ref{eqn:fac_A1A2}) is a more general model where there are two sets of underlying hubs, and the directional relationships are hypothesized to hold from hubs of one kind to hubs of the other kind.
Figure \ref{fig:AA_A1A2_network_compare} illustrate the differences between
the two models.
In the international trade example, Model~(\ref{eqn:fac_AA}) would identify a single set of hubs, corresponding to nodes \#1 to \#4 in the left network plot
in Figure \ref{fig:AA_A1A2_network_compare},
and provide matrices $\bF_t$ that describe how much each hub tends
 to trade with each of the other hubs, corresponding to the colored solid lines connecting different nodes in the figure. A single loading matrix $\bA$ characterizes the relationship between individual countries and the
latent hubs, shown as the green dotted lines connecting countries and hubs. In contrast, Model~(\ref{eqn:fac_A1A2}) provides two sets of underlying hubs: $\bA_1$ relates the countries in their row position to the exporting hubs, corresponding to `Ex' nodes \#1 -- \#4 in the right network plot in Figure \ref{fig:AA_A1A2_network_compare}; and $\bA_2$ relates to the countries in their column position to the importing hubs, corresponding to `Im' nodes \#1 to \#4. The $\bF_t$ then gives the directed relationships from the exporting hubs to the importing hubs. In the international trade example, $\bA_1$ describes countries'
contribution to the exporting hubs and $\bA_2$ countries'
contribution to the importing hubs.

When $\bA_1$ and $\bA_2$ are not linear transformation of one another, Models (\ref{eqn:fac_AA}) and (\ref{eqn:fac_A1A2}) are not equivalent. Consequently, Model~(\ref{eqn:fac_AA}) makes a strong claim about a given data set. When the rows and the columns of a given directional relationship matrix can be demonstrated to span the same space, this agreement is a fact unlikely to arise by chance and probably demonstrates the validity of (\ref{eqn:fac_AA}). With data containing noise, the row and column spaces will probably not match exactly, but a close agreement might still be interpreted as surprising and interesting. However, we will not discuss statistical goodness of fit tests of these two models in this article, but in Section \ref{sec:application} we will demonstrate
detailed comparisons of the two models applied to
the international trade data.


\section{Estimation Procedure and Properties}  \label{sec:est}

Similar to all factor models, the latent factors in the proposed Model~(\ref{eqn:fac_AA}) for asymmetric directional matrix time series can be linearly transformed into alternative but equivalent factors. In general, if $\bH$ is any nonsingular $r \times r$ transformation matrix, we can define an alternative $\bA$ matrix, $\bA^{\ast}$, by letting $\bA^{\ast} = \bA \bH$ and defining the associated $\bF_t$ matrix $\bF_t^{\ast} = \bH^{-1} \bF_t \bH^{' -1}$. Here, we may assume that the columns of $\bA$ are orthonormal, that is, $\bA'\bA=\bI_r$, where $\bI_r$ denotes the identity matrix of dimension $r$. Even with these constraints, $\bA$ and $\bF_t$ are not uniquely determined in (\ref{eqn:fac_AA}), as aforementioned linear transformation is still valid for any orthonormal $\bH$. However, the column space of the loading matrix $\bA$ is uniquely determined. Hence, in what follows, we will focus on the estimation of the column space of $\bA$. We denote the factor loading spaces by $\calM(\bA)$. For simplicity, we will depress the matrix column space notation and use the matrix notation directly.

To facilitate the estimation, we use the QR decomposition $\bA=\bQ \bW$ to normalize the loading matrices, so that Model~(\ref{eqn:fac_AA}) can be re-expressed as
\begin{equation}  \label{eqn:fac_AA_qr}
\bX_t = \bA \bF_t \bA' + \bE_t = \bQ \bZ_t \bQ' + \bE_t, \qquad t = 1, 2, \ldots, T,
\end{equation}
where $\bZ_t = \bW \bF_t \bW'$ and $\bQ' \bQ = \bI_{r}$.

Consider column vectors in (\ref{eqn:fac_AA_qr}), we write
\begin{equation}  \label{eqn:fac_AA_qr_col}
X_{t, \cdot j} = \bA \bF_t A_{j \cdot} + E_{t, \cdot j} = \bQ \bZ_t Q_{j \cdot} + E_{t, \cdot j}, \qquad j = 1, 2, \ldots, n, \;  t = 1, 2, \ldots, T.
\end{equation}

We assume that $\bE_t$ is zero mean. Let $h$ be a positive integer. For $i,j = 1, 2, \ldots, n$, define
\begin{align}
& \bOmega_{zq, ij}(h) = \frac{1}{T-h} \sum_{t=1}^{T-h} \E{\bZ_t Q_{i \cdot} Q^T_{j \cdot} \bZ^T_{t+h}}     \label{eqn:Omega_zqijh} \\
& \bOmega_{x, ij}(h) = \frac{1}{T-h} \sum_{t=1}^{T-h} \E{X_{t, \cdot i} X^T_{t+h, \cdot j}}, \label{eqn:Omega_xijh}
\end{align}
which can be interpreted as the auto-cross-moment matrices at lag $h$ between column $i$ and column $j$ of $\{\bZ_t \bQ'\}_{t=1, \cdots, T}$ and $\{\bX_t\}_{t=1, \cdots, T}$, respectively.

For $h \ge 1$, it follows from (\ref{eqn:fac_AA_qr_col}), (\ref{eqn:Omega_zqijh}) and (\ref{eqn:Omega_xijh}) that
\begin{equation}  \label{eqn:Omega_xijh_Omega_zqijh_relation}
\bOmega_{x, ij}(h) = \bQ \, \bOmega_{zq, ij}(h) \, \bQ'.
\end{equation}

For a predetermined $h_0 \ge 1$, we define
\begin{equation}  \label{eqn:M_col_def}
\bM_{col} = \sum_{h=1}^{h_0} \sum_{i=1}^{n} \sum_{j=1}^{n} \bOmega_{x,ij}(h) \bOmega_{x,ij}(h)' = \bQ \left\{ \sum_{h=1}^{h_0} \sum_{i=1}^{n} \sum_{j=1}^{n} \bOmega_{zq, ij}(h) \bOmega_{zq, ij}(h)' \right\} \bQ'.
\end{equation}

Similar to the column vector version, we define $\bM$ matrix for the row vectors of $\bX_t$'s as following
\begin{equation}  \label{eqn:M_row_def}
\bM_{row} = \sum_{h=1}^{h_0} \sum_{i=1}^{n} \sum_{j=1}^{n} \bOmega_{x',ij}(h) \bOmega_{x',ij}(h)' = \bQ \left\{ \sum_{h=1}^{h_0} \sum_{i=1}^{n} \sum_{j=1}^{n} \bOmega_{z'q, ij}(h) \bOmega_{z'q, ij}(h)' \right\} \bQ',
\end{equation}
where $\bOmega_{z'q, ij}(h) = \frac{1}{T-h} \sum_{t=1}^{T-h} Cov(\bZ'_t Q_{i \cdot}, \bZ'_{t+h} Q_{j \cdot})$ and $\bOmega_{x', ij}(h) = \frac{1}{T-h} \sum_{t=1}^{T-h} Cov(X_{t, i \cdot}, X_{t, j \cdot})$.

Finally, we define $\bM  =  \bM_{col} + \bM_{row}$, that is
\begin{eqnarray}
\bM
& = & \bQ \left\{ \sum_{h=1}^{h_0} \sum_{i=1}^{n} \sum_{j=1}^{n}  \Big[ \bOmega_{zq, ij}(h) \bOmega_{zq, ij}(h)' + \bOmega_{z'q, ij}(h) \bOmega_{z'q, ij}(h)' \Big] \right\} \bQ'.  \label{eqn:M_def}
\end{eqnarray}

Obviously $\bM$ is a $n \times n$ non-negative definite matrix. By Condition 2 and others in \cite{wang2018factor}, it can be shown by similar argument that the right side of (\ref{eqn:M_def}) constitutes a positive definite matrix sandwiched by $\bQ$ and $\bQ'$. Applying the spectral decomposition to $\bM$, we have
\begin{equation}  
\bM  = \bQ \mathbf{U} \mathbf{D} \mathbf{U}' \bQ',  \nonumber
\end{equation}
where $\mathbf{U}$ is a $r \times r$ orthogonal matrix and $\mathbf{D}$ is a diagonal matrix with diagonal elements in descending order. As $\mathbf{U}'\bQ'\bQ\mathbf{U} = \bI_{r}$, the columns of $\bQ\mathbf{U}$ are the eigenvectors of $\bM$ corresponding to its $r$ non-zero eigenvalues. Thus the eigenspace of $\bM$ is the same as $\mathcal{M}(\bQ \mathbf{U})$ which is the same as $\mathcal{M}(\bQ)$.
Under certain regularity conditions, the matrix $\bM$ has rank $r$. Hence, the columns of the factor loading matrix $\bQ$ can be estimated by the $r$ orthogonal eigenvectors of the matrix $\bM$ corresponding to its $r$ non-zero eigenvalues and the columns are arranged such that the corresponding eigenvalues are in the descending order.

Now we define the sample versions of these quantities and introduce the estimation procedure.
For a prescribed positive integer $h_0 \ge 1$, let
\begin{equation}  \label{eqn:Mhat_def}
\hat{\bM} = \sum_{h=1}^{h_0} \sum_{i=1}^{n} \sum_{j=1}^{n} \Big[ \hat{\bOmega}_{x,ij}(h) \hat{\bOmega}_{x,ij}(h)' + \hat{\bOmega}_{x',ij}(h) \hat{\bOmega}_{x',ij}(h)' \Big],
\end{equation}
where $\hat{\bOmega}_{x, ij}(h) = \frac{1}{T-h} \sum_{t=1}^{T-h} X_{t, \cdot i} X'_{t+h, \cdot j}$ and $\hat{\bOmega}_{x', ij}(h) = \frac{1}{T-h} \sum_{t=1}^{T-h} X_{t, i \cdot} X'_{t+h, j \cdot}$. Note that the above calculations are carried out by omitting the NA values. Since the diagonal of the transport volume matrix $\bX_t$ is undefined (NA), omitting the NA's is equivalent to setting them to zero.

A natural estimator for the $\bQ$ specified above is defined as $\hat{\bQ} = \left\{ \hat{\mathbf{q}}, \cdots, \hat{\mathbf{q}}_{r} \right\}$, where $\hat{\mathbf{q}}_i$ is the eigenvector of $\hat{\bM}$ corresponding to its $i$-th largest eigenvalue. Consequently, we estimate the factors and residuals respectively by
\begin{equation}
\hat{\bZ}_t = \hat{\bQ}' \bX_t \hat{\bQ}, \quad \text{and} \quad \hat{\bE}_t = \bX_t - \hat{\bQ} \hat{\bZ}_t \hat{\bQ}' = (\bI_{n} - \hat{\bQ} \hat{\bQ}') \bX_t + \hat{\bQ} \hat{\bQ}' \bX_t (\bI_{n} - \hat{\bQ} \hat{\bQ}').
\end{equation}

The above estimation procedure assumes the number of row factors $r$ is known. To determine $r$ we could use: (a) the eigenvalue ratio-based estimator in \cite{lam2012factor}; (b) the Scree plot which is standard in principal component analysis.  Let $\hat{\lambda}_1 \ge \hat{\lambda}_2 \ge \cdots \ge \hat{\lambda}_{r} \ge 0$ be the ordered eigenvalues of $\hat{\bM}$. The ratio-based estimator for $r$ is defined as
\begin{equation}  \label{eqn:eigen_ratio_r}
\hat{r} = \arg \min_{1 \le j \le r_{\max}} \frac{\hat{\lambda}_{j+1}}{\hat{\lambda}_j},
\end{equation}
where $r \le r_{\max} \le n$ is an integer. In practice we may take $r_{\max} = \lceil n/2 \rceil$ or $r_{\max} = \lceil n/3 \rceil$.

The theoretic properties of the above estimators can be derived trivially from those of the general matrix factor models. For more details, see \cite{wang2018factor}.


\section{Analysis of the International Trade Flow Data}  \label{sec:application}

By examining the network of international trade, we will analyze how countries compare to each other in terms of trade volumes and patterns and how these volumes and patterns evolve as economical cycles and political events unfold.
We want to emphasize that our analysis does not draw on aggregate country statistics such as GNP, production statistics or any other national attributes.


\subsection{Five-Year Rolling Estimation} \label{subsec:5_year_rolling}

To allow for structural changes over time, we break the $408$-month period into $30$ rolling $5$-year periods: $1982$ through $1986$, $1983$ through $1987$ and so forth.
For each $5$-year period, we assume that the loadings are constant and we estimate the loading matrix $\bA$ under Model~(\ref{eqn:fac_AA}) and $\bA_1$ and $\bA_2$ under Model~(\ref{eqn:fac_A1A2}). We estimate three $24 \times r$ loading matrices $\bA$, $\bA_1$ and $\bA_2$ with the same number of factors $r$ across the $30$ periods for comparison purpose. We index these matrices by the mid-year of the five-year periods.

As noted in Section \ref{sec:est}, we can only identify the column spaces of the loading matrices because of the rotational indeterminacy. Let $\bA$ be a matrix whose columns constitute a set of basis of the loading space, then any $\bA \bH$ can be used to represent the column spaces of the loading matrices for any non-singular $r \times r$ matrix. The indeterminacy actually provides flexibility for better model interpretation. Which rotation we select can depend on which perspective we wish to take toward the interpretation of $\bA$ and $\bF_t$. Although in general we might like to seek some kind of approximate simple structure for the columns of $\bA$, this can be done in different ways, corresponding to different orthogonal or oblique rotation criteria in factor analysis.

In the analyses presented in this article, we will adopt as standard a procedure which applies Varimax \citep{kaiser1958varimax} to the columns of $\bA$ \textit{after} they have been scaled to have unit length (eigenvectors are automatically of unit length); this keeps the columns of $\bA$ mutually orthonormal. We further standardize the columns of $\bA$ so that they sum to one. This is feasible because we are dealing with data which contain all positive values, and the estimated columns of $\bA$ contain mostly positive entries with only few negative and small values. It is safe to truncate the negative values to zero while maintain consistency in our estimation. We note that non-negative matrix decomposition can be employed further to make $\bA$ with all positive entries.

When the columns of $\hat{\bA}$
are standardized to sum to one
(i.e. $\bone'\hat{\bA}=\bone'$),
the factor matrix $\bF_t$ can be thought of as a compressed or miniature version of the original observation matrix $\bX_t$. Note that $\E{\bE_t} = 0$, hence $\E{\bone' \bX_t \bone}=\E{\bone' \hat{\bA} \bF_t \hat{\bA}' \bone}=\E{\bone' \bF_t \bone}$. The sum of all the elements in $\bF_t$ is equal to the sum of all elements in $\hat{\bX}_t$, the signal
part of $\bX_t$ fit by the model. The factor matrix $\bF_t$ can be interpreted as expressing relationships among the latent hubs in the same units as the original data. That is, the factor matrix $\bF_t$ can be interpreted as one of the same kind as the original data matrix $\bX_t$, but describing the relations among the latent hubs of the countries, rather than the countries themselves. The diagonals for the observed relational matrices $\bX_t$ are undefined, and will be ignored in the analysis by setting their values to zero. The diagonals for the latent factor matrices $\bF_t$ can be interpreted as the relationship within the same hub, e.g. the import-export between European countries.
With the normalization, the columns of $\hat{\bA}$
show the percentage of contribution each country is to the hub
(from hub's point of view).
This interpretation of our model is different from that of the mixed
membership model \citep{airoldi2008mixed, xing2010state}, where the rows of the membership
matrix sum to one, measuring each actor's percentage of membership to
different communities.




\subsection{Model Trading Volume with Same Export and Import Loadings}  \label{subsec:level_AA}


We first apply Model~(\ref{eqn:fac_AA}) to the international trade volume data. We use the ratio-based method in (\ref{eqn:eigen_ratio_r}) as well as scree plot to estimate the number of latent dimensions. The comparison between these two methods of estimating latent dimensions in different time periods is shown in Table \ref{table:lv_AA_compK}. The scree plot method selects the minimal number of dimension that explain at least 85 percents of the variance in the original data. The estimate by (\ref{eqn:eigen_ratio_r}) tends to be smaller than
the one given by scree plot. The percentage of total variance explained by the $r=4$ factor model is shown in the last line.

\begin{table}[htpb!]
\centering
\resizebox{\textwidth}{!}{%
\begin{tabular}{c|ccccccccccccccc}
\hline
 & 1984 & 1985 & 1986 & 1987 & 1988 & 1989 & 1990 & 1991 & 1992 & 1993 & 1994 & 1995 & 1996 & 1997 & 1998 \\ \hline
Ratio  & 2 & 4 & 1 & 1 & 1 & 1 & 5 & 1 & 2 & 2 & 2 & 2 & 2 & 2 & 2 \\
Scree  & 2 & 3 & 4 & 4 & 4 & 5 & 5 & 5 & 4 & 3 & 3 & 3 & 3 & 2 & 2 \\ \hline
$r=4$  & 97 & 94 & 91 & 89 & 85 & 83 & 83 & 84 & 88 & 91 & 90 & 91 & 93 & 94 & 94 \\ \hline \hline
 & 1999 & 2000 & 2001 & 2002 & 2003 & 2004 & 2005 & 2006 & 2007 & 2008 & 2009 & 2010 & 2011 & 2012 & 2013  \\ \hline
Ratio & 2 & 2 & 2 & 2 & 2 & 2 & 2 & 2 & 6 & 5 & 2 & 2 & 2 & 2 & 2   \\
Scree & 3 & 4 & 3 & 4 & 3 & 3 & 3 & 4 & 4 & 4 & 4 & 4 & 3 & 3 & 4   \\ \hline
$r=4$ & 90 & 89 & 91 & 91 & 91 & 91 & 90 & 88 & 86 & 86 & 88 & 90 & 92 & 92 & 88  \\ \hline
\end{tabular}%
}
\caption{Estimated latent dimension of $\bF_t$ in Model~(\ref{eqn:fac_AA}) between ratio-based and scree plot methods. 
	The last line presents the percentage of total variance explained by the $r=4$ factor model.}
\label{table:lv_AA_compK}
\end{table}

As shown in Table \ref{table:lv_AA_compK}, most dimension estimates are
smaller than or equal to 4
and the factor model with $r=4$ explains at least $83\%$ of the total variance. Thus, latent dimension $r=4$ will be used for illustration in all periods for ease of comparison. We will focus on the loading matrix $\bA$, which prescribes the interpretations of the latent hubs by linking them to the observed countries, and the factor matrix $\bF_t$, which characterizes the directional relationship between latent hubs.

\begin{figure}[htpb!]
	\centering
	\includegraphics[width=\textwidth, keepaspectratio]{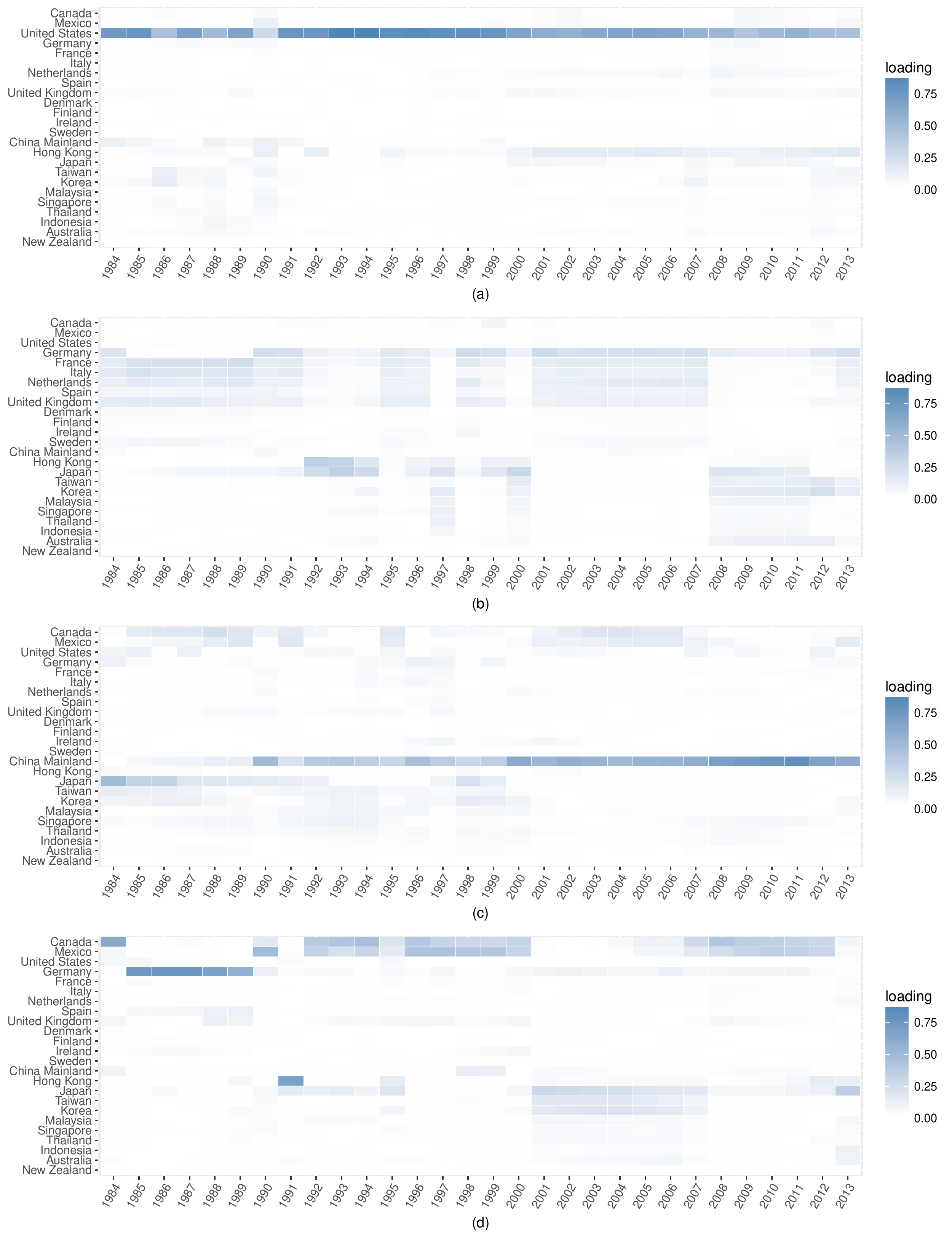}
	\caption{Latent factor loadings for trading volume on $r = 4$ hubs for a series of $30$ rolling five-year periods indexed from $1984$ to $2013$.}
	\label{fig:lv_AA_dyn_4_fac_all}
\end{figure}

Figure \ref{fig:lv_AA_dyn_4_fac_all} presents the heat maps of the loadings
of each country/region
on the top four latent hubs from $1984$ to $2013$. See supplementary material for plotted values. Four vertically aligned heat maps correspond to four columns of loading matrix $\hat{\bA}$ from year $1984$ to $2013$. For example, the first columns (denoted by 1984) of the plot (a), (b), (c), and (d) are the four columns of the loading matrix $\hat{\bA}_{1984}$ calculated using data from 1982 to 1986; the second columns (denoted by 1985) of the four heat maps correspond to the four columns of the loading matrix $\hat{\bA}_{1985}$ calculated using data from 1983 to 1987; and so on.


Although traditional eigen-analysis arranges spectral decomposition using the rank of eigen-values, the choice of ranking is actually flexible. We choose to rank the columns of $\hat{\bA}$ from different years according to their maximum loading on the United States, United Kingdom, and China Mainland for plots (a), (b) and (c).
The reason for our choice is that the structure of international trade changes over time. The latent factors or hubs may rank differently in terms of their accounted variances at different time periods. For example, latent hub of European countries
accounts for the largest portion of variance in 1985, but it
ranks the third
in 2001 and even no longer belongs to the top four hubs in 2009.
Plot (d) contains the remaining factor for all the years. In such representation, plots (a), (b), (c) and (d) are considered together as top four hubs.
The factors in one heat map may ranked differently in terms of accounted variance at different times. But they correspond to the same interpretation at certain time periods.

Recall that each column in a heat map sums up to one. Thus, the value at each cell reflects a country's contribution in the corresponding hub at a certain year. For example in Figure \ref{fig:lv_AA_dyn_4_fac_all} (a), the darkest cell corresponds to USA at year 1984 indicates that the portion of trading taken by USA on latent hub (a) is the largest among all countries. The changes of color intensity of the cells shows the evolution in a country's participation in the four hubs over $30$ years.

The latent hub corresponding to Figure \ref{fig:lv_AA_dyn_4_fac_all} (a) can be interpreted as a United States dominated hub, as the loadings of the United States on this hub are much larger than that of all other countries. From the plot, it is clear that the United States dominates this hub very strongly from $1984$ to $2013$. However, its contribution gradually decreases since $2002$ and reaches its minimal from year $2009$ onwards, possibly due to the aftermath of the $2008$ financial crisis. The decrease from United States is offset by the increase from United Kingdom, Netherlands, Hong Kong, Japan, Taiwan and Korea, which is manifested by the increasingly darker cells since $2002$ for those countries in Figure (a).

The latent hub corresponding to Figure \ref{fig:lv_AA_dyn_4_fac_all} (b) are aligned according to the maximum loading on United Kingdom, and not surprisingly, it is also heavily loaded by European countries such as France, Italy, Netherlands, Spain and Germany. Therefore, this hub can be interpreted as a hub dominated by European countries. From 1985 to 1989, Germany's trading was so distinctive from other European countries that it took a separate hub as shown in Figure \ref{fig:lv_AA_dyn_4_fac_all} (d). During this period, France, United Kingdom, Italy and Netherlands accounted for a large portion of European's trading. After 1990, Germany, France, United Kingdom, and Italy took approximately equal portions. With the introduction of Euro in $2002$, Netherlands, Spain, and United Kingdom's contributions in trade increase. We also note that the loading of some Asian economies, such as Hong Kong, Japan, Taiwan, Malaysia and Singapore, on this hub is also significant in certain periods including from 1992 to 1994, and from 2008 onwards. This suggests that, in these periods, the hub representing Asian economies explain more variance in the original data than the European hub and replace European hub as one of the top four hubs. 

The latent hub corresponding to Figure \ref{fig:lv_AA_dyn_4_fac_all} (c) are hubs that China Mainland has maximum loadings on. Before 1989, Japan loads more on this hub than China Mainland does. China Mainland's loading on this hub kept increasing throughout the period. Its contribution to the hub becomes larger than Japan's  from the year 1989. It shows a clearer transition of trading centrality of large Asia economies, though Japan is also actively participating in all other hubs.

The latent hub corresponding to Figure \ref{fig:lv_AA_dyn_4_fac_all} (d) features sizable loadings on Canada, Mexico, Japan, Taiwan, and Korea. Thus the fourth hub of the latent factor matrix represents the group of large economies in North American and Asia-pacific except for the US and China Mainland. The evolution of the hub (d) is striking. Before $1989$, Germany's trading is so distinctive from the other European countries that it uses this single hub exclusively. After that, this hub is dominated by NAFTA countries between 1990 and 2000 and between 2007 and 2012. APEC countries dominated this hub between 2001 and 2007.

\begin{figure}[htpb!]
	\centering
	\includegraphics[width=\textwidth, keepaspectratio]{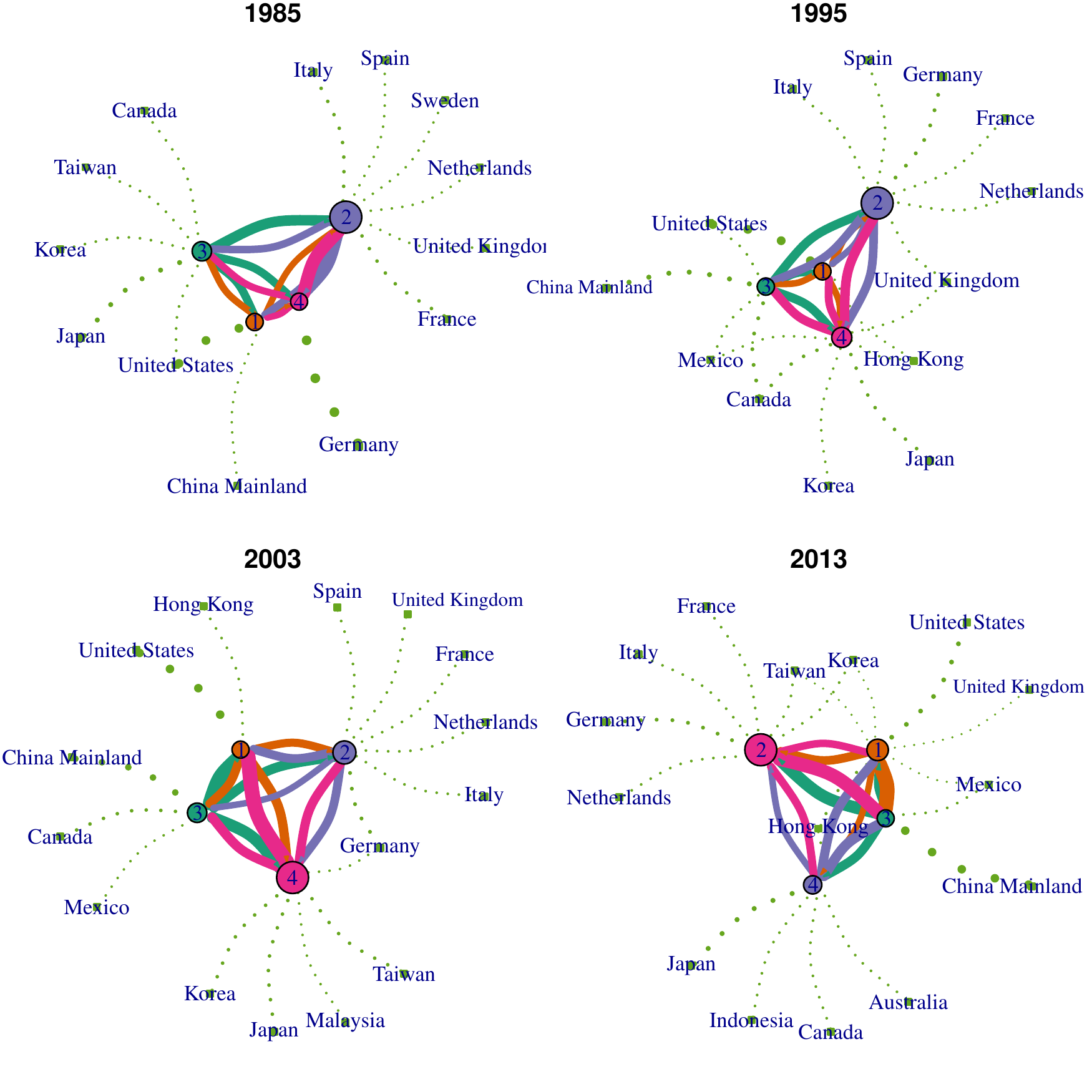}
	\caption{Trading volume network plot of latent hubs and relationship between countries/regions and the latent hubs. Thickness of the solid line reflects the volume of trades among latent hubs. Thickness of the dotted lines represents the level of  contribution. Note that a country can have significant contributions to multiple latent hubs after truncation. }
	\label{fig:lv_AA_network_rr_4_combine}
\end{figure}

Figure \ref{fig:lv_AA_network_rr_4_combine} plots the trading network among four latent hubs as well as the relationship between countries and latent hubs for four selected years. The trading network among latent hubs is plotted based on the average of $4 \times 4$ latent factor matrix $\bF_t$ in the corresponding 5-year rolling window. The colored circles represent four latent hubs. Note that the eigen-decomposition algorithm we used does not guarantee positive entries in $\bF_t$ (hubs). The negative values in $\bF_t$ are interpreted as a change of trading direction. Non-negative matrix factorization proposed by \cite{lee2001algorithms} can be used to ensure positive entries, though we did not use it here for simplicity. There are very few negative entries in this example. The size of each circle conveys the trading volumes within the hub, i.e., the values of the diagonal elements in the latent factor matrix. The width of the solid lines connecting the circles conveys the trading volume between different hubs, i.e., the values of the off-diagonal elements in the latent factor matrix. The direction of the flow is conveyed by the color of the line. Specifically, the color of the line is the same as its export hub. For example, a blue line connecting a blue node and a red node represents the trade flow from the blue node to the red node. Note that the widths of the solid lines across different network plots are not comparable because they are scaled to fit each individual plot for different years because the trading volume changes dramatically in the period.

The relationships between countries and the four hubs, shown as the dotted lines, are plotted using a truncated version of the estimated loading matrix $\hat{\bA}$ to provide an uncluttered view that only captures the prominent relations. The truncation is achieved by first rounding all entries of $10 \hat{\bA}$ to integers and then normalizing the non-zero entries to have column sum one.

Clearly shown in the network plot, in 1985
the United States and Germany solely dominate hubs \#1 and \#4,
respectively. Latent hub \#2 is mainly used by
European countries such as Spain, Netherlands, France, Sweden, United Kingdom and Italy.
Latent hub \#3 is mainly used by
Japan, Korea, Taiwan and Canada. As shown by the thick orange lines, hub \#1, representing the U.S., exports mostly to hub \#3, which load mostly on large Asian economies and Canada. The thick pink and purple lines connecting hubs \#4 and \#2 imply that Germany trades a lot with other European countries even through itself stands out from the European countries.

In 1995, European countries become closer and they all mainly use
 a single hub \#2, which reflects the effects of the foundation of European Union in 1993. The trade within the hub (group)
is the largest, indicating the strong inter-European trading activities.
The year of 1995 also celebrates developments of Asian countries when they dominate two latent hubs, namely hub \#3 and \#4. This can be explained by the fast development of these Asian countries to emulate the developed economies in North American and European economies during the late 80's and early 90's. There are large amount of exporting from Asian countries to the United States and European countries as indicated by the thick pink and green lines to hub \#1 and hub \#2. Also, the trading among Asian countries is also large as shown by the thick lines connection green hub \#3 and pink hub \#4. Mexican and Canada also contribute to these two hubs.

In 2003, hub \#3 is mostly used by Canada, China Mainland and Mexico. It represents the latent hub that exports a lot to hub \#1 (Hong Kong and United States). Hub \#2 that contributed meaningfully by Netherlands, France, Italy, Spain, United Kingdom stays the same as the European hub in 1995. Hub \#4 can be interpreted as APEC hub because it is mostly composed of Japan, Taiwan, Korea, and Malaysia. The United States still loaded on hub \#1. However, Hong Kong also load heavily on this hub, indicating that these two countries share some similar import/export patterns.
For example, Hong Kong trades a lot with
 Canada, China Mainland and Mexico (the thick orange and green lines between nodes \#1 and \#3) and it also imports a large volume from the APEC type \#4. The exporting volumes from hub \#3 (Canada, China Mainland and Mexico) to U.S. hub \#1 and from APEC hub \#4 to U.S. hub \#1 are among the largest trading volumes in this period.

In 2013, China Mainland dominates in a single hub \#3, indicating China Mainland's growing importance in international trade in the 2010's. Netherlands, Germany, Italy and France remain contributing a large portion to the European hub \#2, while United Kingdom contribute more to the
 United States hub \#1. The United States still loads completely on hub \#1. At the same time, the compositions of the hubs are less geographically concentrated -- European hub \#2 and United States hub \#1 are shared with Taiwan, Korea, Mexico and Hong Kong and Asian hub \#4 is shared with Canada -- which indicates that international trades are more global and less regional in the 2010's.

\begin{figure}[htpb!]
	\centering
	\includegraphics[width=\textwidth, keepaspectratio]{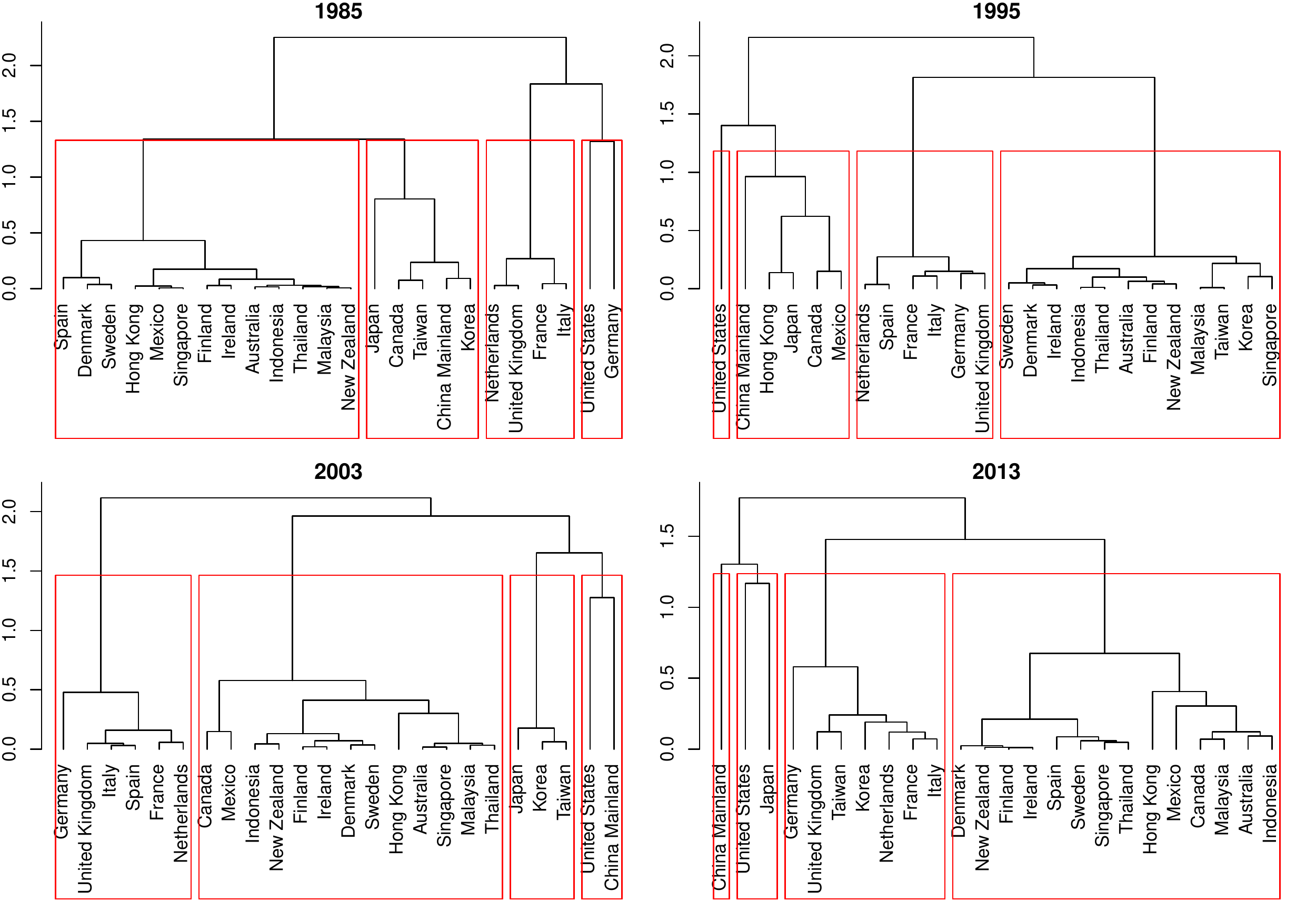}
	\caption{Clustering of countries based on their loading.}
	\label{fig:lv_AA_dendrogram_rr_4_group_4_combine}
\end{figure}

A hierarchical clustering algorithm \citep{xu2005survey, murtagh2014ward} is employed to cluster countries based on their contribution patterns over years under Euclidean distance and the {\it ward.D} criterion.
The dendrograms in Figure \ref{fig:lv_AA_dendrogram_rr_4_group_4_combine} shows detailed structures of the hierarchical clustering results. The rectangles denote clusters that divide countries into four groups. It offers a
different perspective to inspect the dynamics of countries' trading behaviors. Generally speaking, geographically or culturally proximate countries are usually in the same group and behave similarly. For example, one can easily identify the European group and the Asia-pacific group from the dendrograms. Countries with similar trading behaviors also tend to be clustered in
the same group. For example in the 1990's, Canada and
Mexico are in the same group with Hong Kong, Japan and
China Mainland -- they all export in large volumes to United States. The overall structure of international trading seems steady over years: fours groups in all years can be labeled as `United States', `European active', `Asia-pacific active', `European-Asia-pacific less active'. However, the relationship between individual countries are changing over the period. In the 1980's, United States and Germany are in the same group, reflecting the fact that they are the most active countries in trading, especially exporting, in the 80's. In the 1990's, United States accounts for a single group because of its dominant position in the international trades in the decade. China Mainland's participation in the global trade has been gradually increasing over the years: in 1980's China Mainland's trading behavior is more like economies such as Korea. From 1990's to 2010's, as China Mainland becomes more active in importing and exporting, its trading behavior becomes more similar to that of the United States. Later its trading behavior becomes so distinctive that it makes up single cluster in the 2010's. Again, these patterns resonate to some of the observations from Figures \ref{fig:lv_AA_dyn_4_fac_all} and \ref{fig:lv_AA_network_rr_4_combine}.

\subsection{Model Trading Volume with Different Export and Import Loadings} \label{subsec:level_A1A2}

Now we apply Model~(\ref{eqn:fac_A1A2}) to the
international trade volume data. We use the ratio-based method in (\ref{eqn:eigen_ratio_r}) as well as scree plot to estimate the number of latent dimensions. The comparison between these two methods of estimating importing and exporting dimensions in different time periods is shown in Table \ref{table:lv_A1A2_compK}. Note that Model~(\ref{eqn:fac_A1A2}) assumes different exporting and import loadings $\bA_1$ and $\bA_2$. Similar to Figure \ref{table:lv_AA_compK} , the scree plot method selects the minimal number of dimension that explain at least 85 percent of the variance in the original data. The percentage of total variance explained by the $4 \times 4$ factor model is shown in the last line. With the additional flexibility of allowing different row and column loading matrix, the estimated dimension is slightly smaller than that in Table \ref{table:lv_AA_compK}, though the ratio estimate becomes less stable.

\begin{table}[htpb!]
\centering
\resizebox{\textwidth}{!}{%
\begin{tabular}{c|cccccccccc}
\hline
 & 1984 & 1985 & 1986 & 1987 & 1988 & 1989 & 1990 & 1991 & 1992 & 1993 \\ \hline
Ratio & (1, 1) & (1, 1) & (8, 1) & (1, 1) & (11, 1) & (6, 1) & (6, 3) & (2, 1) & (2, 1) & (2, 2) \\
Scree & (2, 2) & (2, 2) & (3, 3) & (3, 3) & (4, 4) & (5, 4) & (4, 4) & (4, 4) & (3, 3) & (3, 3) \\ \hline
(4,4) & (98, 98) & (95, 96) & (92, 94) & (91, 92) & (85, 91) & (85, 90) & (88, 89) & (91, 90) & (94, 93) & (95, 94) \\ \hline \hline
 & 1994 & 1995 & 1996 & 1997 & 1998 & 1999 & 2000 & 2001 & 2002 & 2003 \\ \hline
Ratio & (5, 2) & (2, 2) & (2, 2) & (2, 2) & (2, 2) & (1, 1) & (1, 1) & (1, 1) & (1, 1) & (1, 1) \\
Scree & (3, 3) & (3, 3) & (2, 2) & (2, 2) & (2, 2) & (3, 3) & (3, 3) & (3, 3) & (3, 3) & (3, 3) \\ \hline
(4,4) & (93, 92) & (95, 94) & (96, 95) & (97, 97) & (96, 96) & (94, 94) & (92, 92) & (93, 94) & (95, 95) & (93, 93) \\ \hline \hline
 & 2004 & 2005 & 2006 & 2007 & 2008 & 2009 & 2010 & 2011 & 2012 & 2013 \\ \hline
Ratio & (1, 1) & (6, 6) & (1, 1) & (6, 6) & (1, 6) & (1, 6) & (1, 5) & (5, 5) & (7, 1) & (1, 1) \\
Scree & (3, 3) & (3, 3) & (3, 3) & (4, 4) & (4, 4) & (4, 3) & (3, 3) & (3, 3) & (3, 3) & (3, 3) \\ \hline
(4,4) & (94, 93) & (93, 93) & (91, 91) & (88, 89) & (88, 91) & (89, 91) & (92, 93) & (94, 94) & (95, 95) & (90, 91) \\ \hline
\end{tabular}%
}
\caption{Comparison of estimated latent dimension of $\bF_t$ in Model~(\ref{eqn:fac_A1A2}) between ratio-based and scree plot methods. 
The last line presents the percentages of variance explained by the $4 \times 4$ factor model in (export, import), respectively.}
\label{table:lv_A1A2_compK}
\end{table}

As shown in Table \ref{table:lv_A1A2_compK}, most dimension estimators are smaller than 4 and the factor model with dimension $4 \times 4$ explains at least $85\%$ of the total variance. Thus, latent dimension $4 \times 4$ will be used for illustration and comparison between different period. In the following analysis, we employ the same visualization tools as those in Section \ref{subsec:level_AA}. However, there are separate plots for loading matrices $\bA_1$ and $\bA_2$ since Model~(\ref{eqn:fac_A1A2}) differentiates the importing and exporting dimensions.

\begin{figure}[htpb!]
	\centering 
	\includegraphics[width=\textwidth, keepaspectratio]{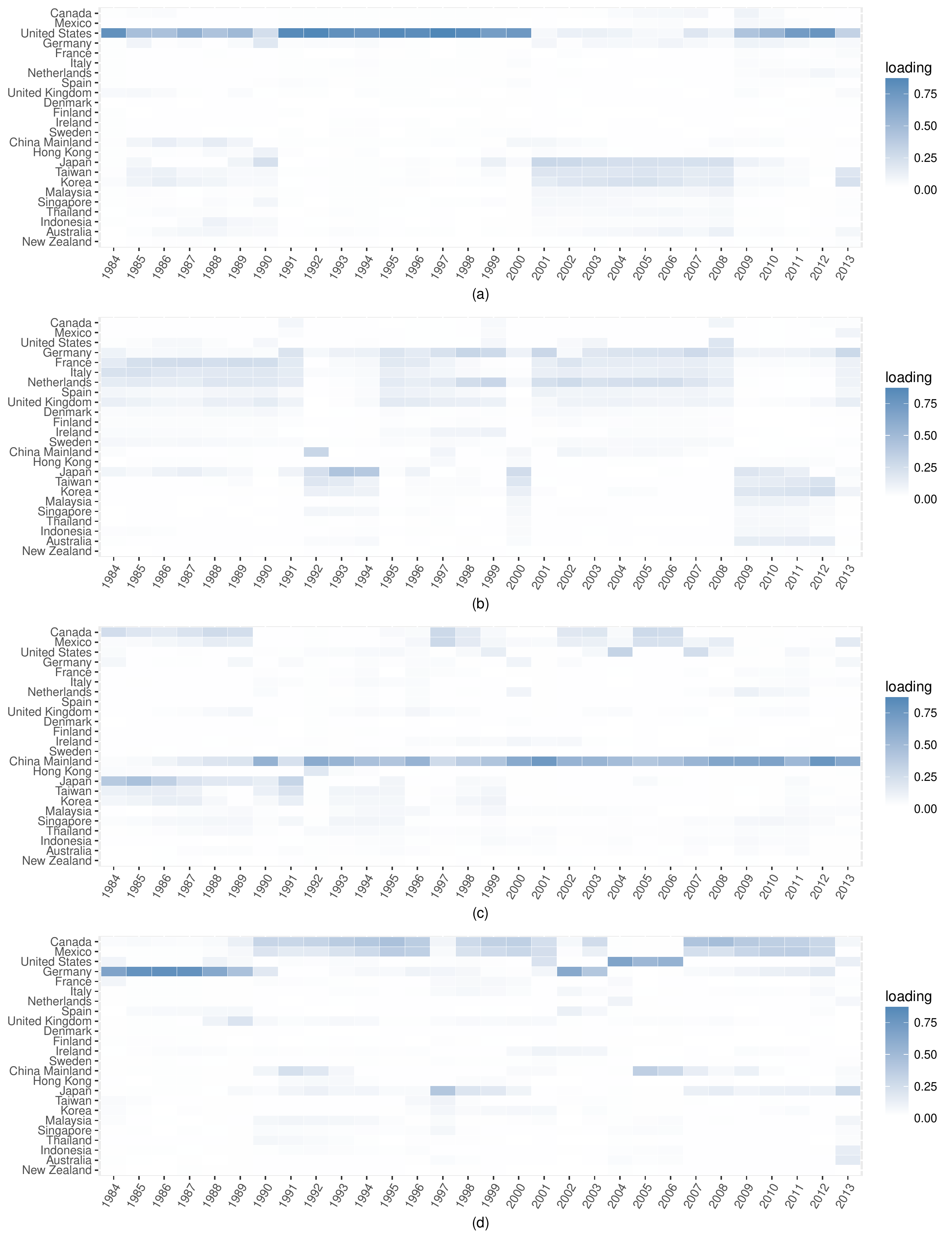}
	\caption{Latent export loadings for trading volume on $r = 4$ hubs for a series of $30$ rolling five-year periods indexed from $1984$ to $2013$. }
	\label{fig:lv_export_dyn_4_fac_all}
\end{figure}

\begin{figure}[htpb!]
	\centering 
	\includegraphics[width=\textwidth, keepaspectratio]{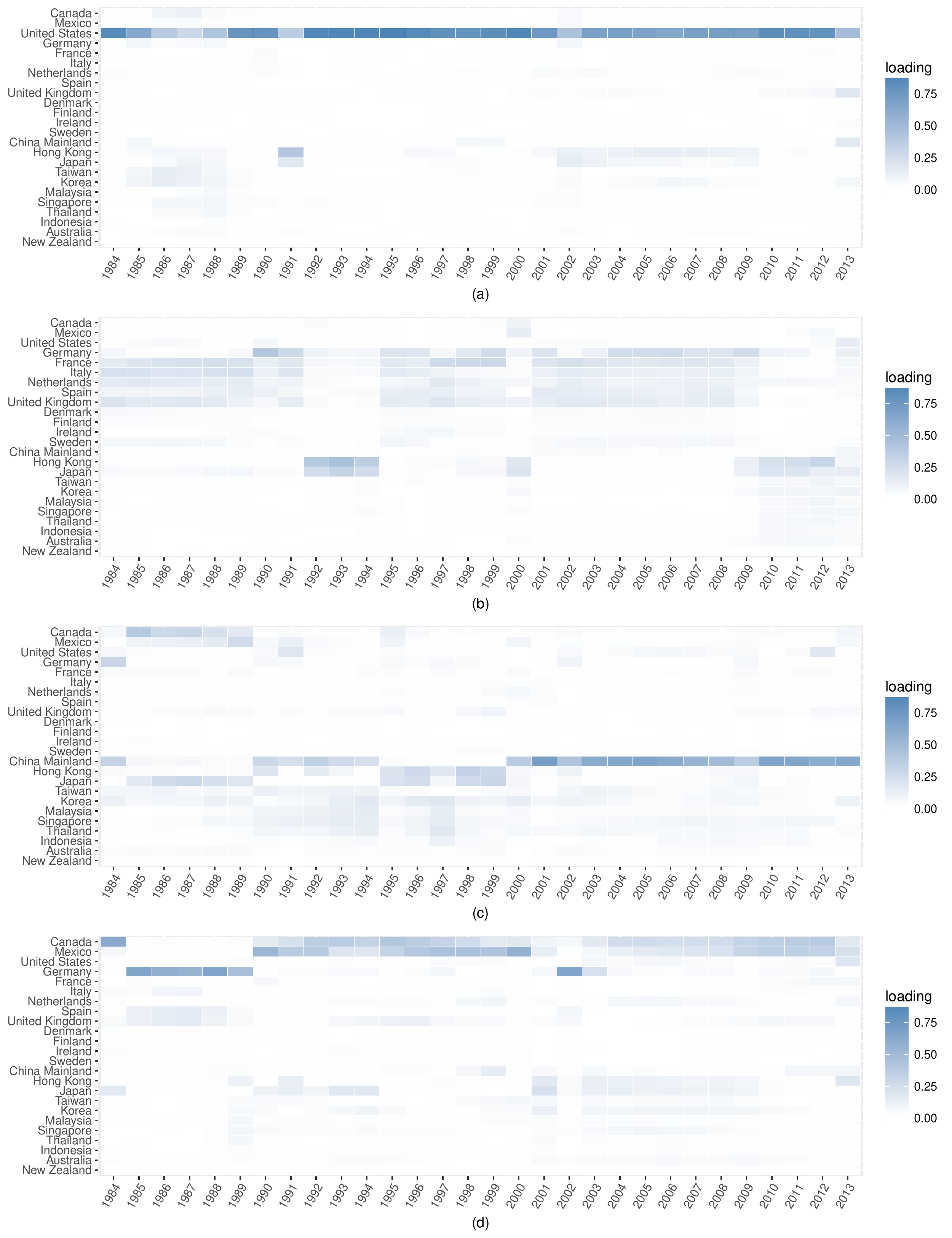}
	\caption{Latent import loadings for trading volume on $r = 4$ hubs for a series of $30$ rolling five-year periods indexed from $1984$ to $2013$. }
	\label{fig:lv_import_dyn_4_fac_all}
\end{figure}

Figures \ref{fig:lv_export_dyn_4_fac_all} and \ref{fig:lv_import_dyn_4_fac_all}
 present the heat maps for exporting loading $\bA_1$ and importing loading $\bA_2$, respectively. They are designed in the same way as those in Figure \ref{fig:lv_AA_dyn_4_fac_all}. The patterns are strikingly similar in the heat maps of $\bA$, $\bA_1$, and $\bA_2$. Plots (a) in all three Figures \ref{fig:lv_AA_dyn_4_fac_all}, \ref{fig:lv_export_dyn_4_fac_all} and \ref{fig:lv_import_dyn_4_fac_all} represent the latent hub of United States. Plots (b), (c), and (d) in all figures represent the latent hub of European countries, Japan/China Mainland, and NAFTA countries (except US), respectively. The loadings of countries on these top four latent hubs evolve in the same way among these three figures.

There are a few
noticeable differences in the import and export behavior though.
For example, US's import activities
dominate the import hub \#1 throughout the period, but its export activities
weaken in the export hub \#1 during the 2000's, facing competition from the
Asian countries. China's export activities start in the early 1990's but
it's import activities only show dominance in the 2000's.

\begin{figure}[htpb!]
	\centering
	\includegraphics[width=\textwidth, keepaspectratio]{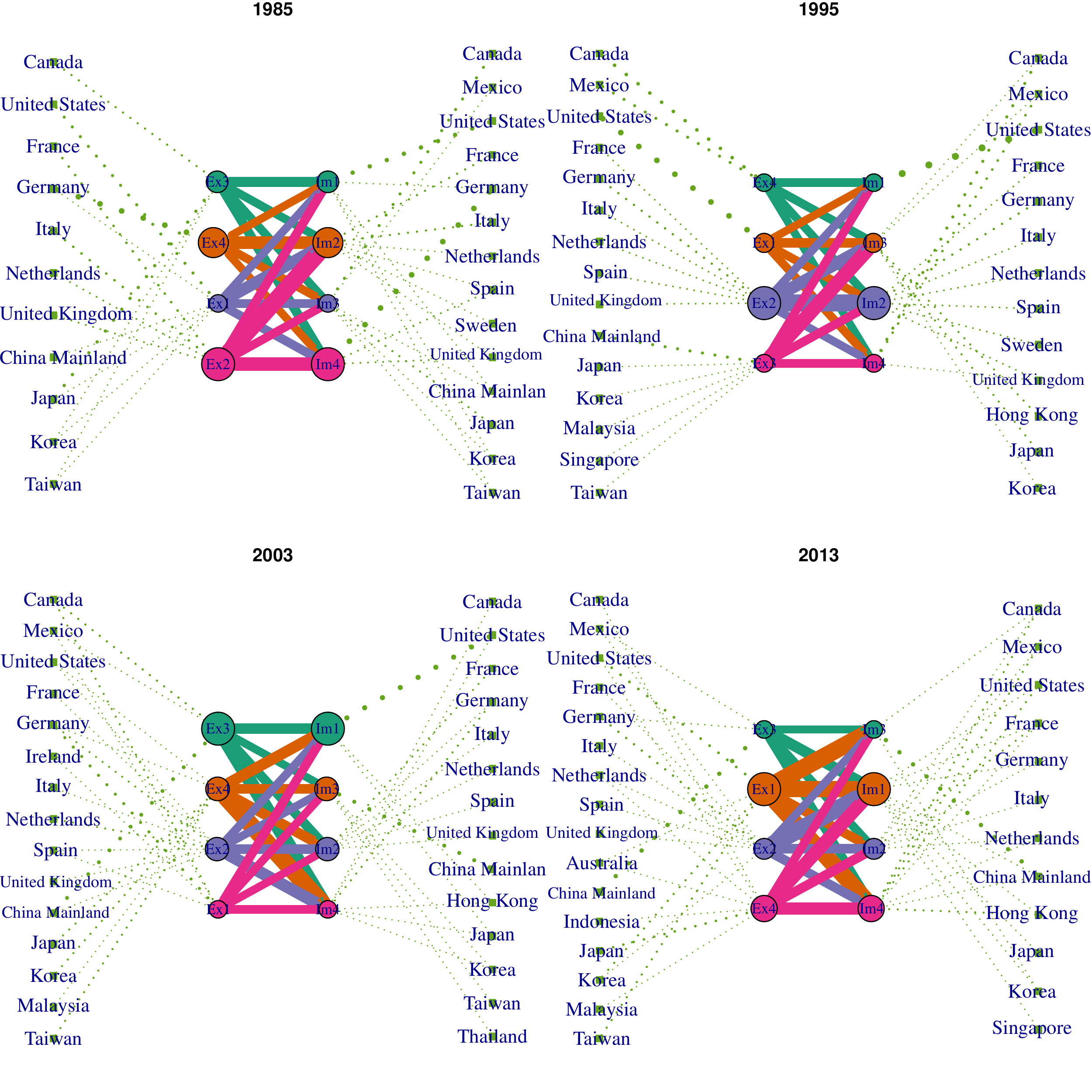}
	\caption{Trading volume network plot of latent hubs and relationship between countries and the latent hubs. Thickness of the solid line represents the volume of trades among latent hubs. Thickness of the dotted lines represents the level of connection between latent hubs and countries. Note that a country can be related to multiple latent hubs. To provide a clear view, $\hat{\bA_1}$, and $\hat{\bA_2}$ are truncated by rounding $10 \hat{\bA}$ and then normalizing the non-zero entries to have column sum one.}
	\label{fig:lv_A1A2_network_rr_4_combine}
\end{figure}

Figure \ref{fig:lv_A1A2_network_rr_4_combine} plots the trading network among four latent hubs as well as the relationship between countries and latent hubs for four selected years. Since we use different export (left) and import (right) loading matrix, the relationships between countries and latent hubs are different for import and export activities. The meanings of row and column dimensions of the latent factor matrix $\bF_t$ are different too. Specifically, the rows of $\bF_t$ represents the exporting hubs while the columns correspond to the importing hubs. Thus we distinguish the row and column hubs and have eight circle nodes for the latent hubs in Figure \ref{fig:lv_A1A2_network_rr_4_combine}. The nodes annotated with ``Ex'' and ``Im'' correspond to the export (row) hubs and the import (column) hubs, respectively. We notice symmetry between the exporting and importing nodes or hubs, indicating empirically the validity of Model~(\ref{eqn:fac_AA}), for certain years. For example in 1995, the exporting node ``Ex1'' and importing node ``Im1'' both represent the United States hub; the exporting node ``Ex2'' and importing node ``Im2'' both represent the Europe hub; and the exporting node ``Ex3'' and importing node ``Im3'' both represent the Asia hub;the exporting node ``Ex4'' and the importing node ``Im4'' both represent the Canada \& Mexico hub. However, in this paper we do not devise a formal statistical method for testing Model~(\ref{eqn:fac_AA}) and (\ref{eqn:fac_A1A2}), which is an important problem for future research.

\begin{figure}[htpb!]
	\centering
	\includegraphics[width=\textwidth, keepaspectratio]{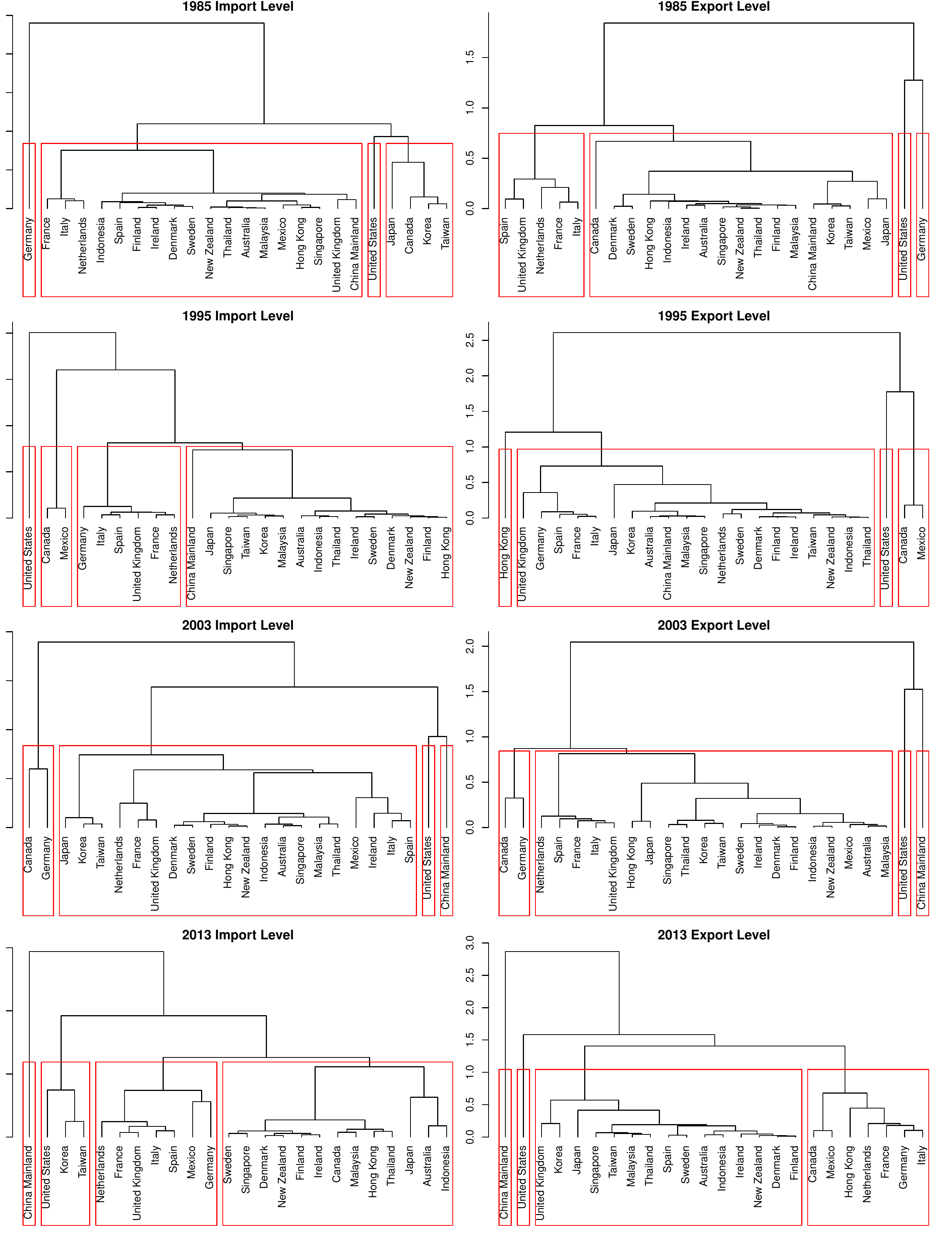}
	\caption{Clustering of countries based on their trading volume latent hub representations.}
	\label{fig:lv_A1A2_dendrogram_combine_rr_4_sel}
\end{figure}

\section{Summary and Conclusion}  \label{sec:summary}

In this paper, we proposed an innovative framework of modeling dynamic transport networks and an effective method to estimate the dynamic structure that underpins the surface networks. We have collected, cleaned, and analyzed a data set of a dynamic transport network of monthly international trade volumes among 24 countries and regions over 34 years. We have investigated the trading hubs, centrality, patterns and trends in the trading network of the
24 countries and regions under the proposed framework and methodology. The results are able to offer sensible insights in international trading and show matching change points to trading policies.

Unlike the traditional node-and-edge level modeling of dynamic networks, which mainly focus on the link connectivity, the framework and the estimation method proposed in this paper offers an effective way for unveiling latent structure of the surface nodes and their relations in a dynamic transport network. The proposed methodology has several distinctive features in its structure and implementation. First, the matrix-variate time series modeling concisely captures the amount of data (traffic) moving across a network. The direction and size of a traffic is captured by the location and value in matrix, respectively; second, we impose neither any distributional assumptions on the underlying network nor any parametric forms on its covariance function. The latent network is learned directly from the data with little subjective input; and third, the idea is simple, yet quite general and flexible. It can be easily extended to include factor dynamics and covariates.


Our results on international trade flow consist of two major parts: the latent factor matrices that capture the structure and the dynamics of the latent low-dimensional network; and the loading matrices that connect the latent nodes with the surface nodes and characterize the semantics of the latent nodes.

Based on the latent factor matrices and the loading matrices, we have the findings on (i)~meaningful \textbf{\textit{trading hubs}} that aggregate and distribute trading flows among countries over the three decades. (ii)~distinct countries that are \textbf{\textit{central}} in the sense that some hubs are used exclusively by them  and the changes of centrality in international trading. (iii)~international trading \textbf{\textit{patterns and trends}} for the 24 countries and regions and for the latent trading hubs. These findings are elaborated in detail in the following paragraphs.
\begin{enumerate}[label=(\roman*)]
\setlength{\itemsep}{0pt}
\item \textbf{\textit{Trading Hubs.}} Figure \ref{fig:lv_AA_dyn_4_fac_all} consists of heatmaps of loading matrices that characterize the connection between surface nodes and latent nodes (trading hubs). Figure \ref{fig:lv_AA_network_rr_4_combine} consists of network plot of the latent nodes and the connection between surface nodes and latent nodes (trading hubs). Both show four major trading hubs, namely, the United States, European countries, large Asian economies, and German in the 1980's and 1990's or North American countries other than United States in the 2000's and 2010's. Figures \ref{fig:lv_export_dyn_4_fac_all}, \ref{fig:lv_import_dyn_4_fac_all}, and \ref{fig:lv_A1A2_network_rr_4_combine} differentiate the exporting and importing behaviors but suggests the same trading hubs.
\item \textbf{\textit{Centrality.}} In the 1980's, United States and Germany are the two largest economies that trade in large volumes and with a wide range of countries. United States keeps its central role from 1980's to 2010's. Germany lost its centrality in the late 1990's. China Mainland gradually grows its trading capacity in the late 1990's and assumes a central role in the 2000's and 2010's. Throughout decades, European has been a tight trading block that trades more with its own group members than
with outside countries. Observations from Figures \ref{fig:lv_export_dyn_4_fac_all}, \ref{fig:lv_import_dyn_4_fac_all}, and \ref{fig:lv_A1A2_network_rr_4_combine} are the same for the evolution of centrality.
\item \textbf{\textit{Patterns and trends.}} Note that the analysis is based on a five-year window rolling analysis. Our estimates are a sequence of time-evolving loading matrices and factor matrices that capture the dynamics of the networks of trading hubs and countries. Figure \ref{fig:lv_AA_dyn_4_fac_all} provides several interesting observations of the pattern changes over the decades. First, the United States uses the first hub exclusively all the time, although its contribution to this hub is slightly decreasing in the 2000's and 2010's. Second, European countries uses the second hub exclusively, though its dominance in this hub was interrupted by emerges of developing Asian countries from 1992 to 1994 and from 2008 to 2011. The first period corresponds to the growth of the Four Asian Tiger economies that is attributed to export oriented policies and strong development policies. The second period corresponds to the 2008 financial crisis that affected most European countries. Third, Japan uses the large Asian economies hub exclusively in the 1980's and China Mainland has gradually taken it
over since 1990's. Forth, in the 1980's Germany is different from the other European countries that itself exclusively use one single hub. Gradually, Germany has blended in the European and become a member sharing the latent European hub since the formation of European Union in 1993. This phenomenon is mostly prominent after common currency ''euro`` was established in 1999 and came into full force in 2002. The fourth latent hub is later taken over by Canada and Mexico who trades heavily with United States. Figure \ref{fig:lv_AA_network_rr_4_combine} in addition offers more detail in the trading between latent trading hubs. Figure \ref{fig:lv_export_dyn_4_fac_all}, \ref{fig:lv_import_dyn_4_fac_all}, and \ref{fig:lv_A1A2_network_rr_4_combine} presents more details in exporting and importing patterns and trends.
\end{enumerate}


There are many directions where we can extend our current work. The proposed methods are able to effectively reduce the dimension of the dynamic networks and uncover its core structure. The estimated latent dynamic networks and its relation with the surface networks can be further used for testing and predicting the networks. Also, current model does not explicitly model the dynamics of matrix factors. Incorporating an autoregressive model for the latent matrix factors will enable prediction of future network flows. This will result in a dynamic factor model for matrix-variate time series. Including covariates of nodes, such as the GDP of the country or geographic distance between countries, will also be interesting future research. Methods for testing the models (\ref{eqn:fac_AA}) and (\ref{eqn:fac_A1A2}) are also of great importance.


\newpage

\newpage
\begin{singlespace}
\bibliographystyle{apalike}
\bibliography{dnet}
\end{singlespace}

\end{document}